\documentclass[11pt,a4paper,hyperref]{article}

\usepackage[utf8]{inputenc}
\usepackage{graphicx}
\usepackage{amsmath}
\usepackage{hyperref}
\allowdisplaybreaks[2]

\newcommand{\chpt}{ChPT}
\newcommand{\Lag}{\mathcal{L}}

\newcommand{\tr}{\mathrm{Tr}}

\begin{document}
\begin{titlepage}
\begin{flushright}
LU TP 16-50\\
September 2016\\
Revised February 2017\\
\end{flushright}

\hfill

\begin{center}
{\Large\bf Twisted finite-volume corrections to $K_{l3}$ decays\\[3mm]
with partially-quenched and rooted-staggered quarks}
\\[2cm]
{\bf Claude Bernard$^a$, Johan Bijnens$^b$, Elvira G\'amiz$^c$
and Johan Relefors$^b$}
\\[1cm]
$^a$Department of Physics, Washington University, St. Louis, Missouri, USA
\\[0.3cm]
$^b$Department of Astronomy and Theoretical Physics, Lund University,
Lund, Sweden
\\[0.3cm]
$^c$CAFPE and Departamento de F\'\i sica Te\'orica y del Cosmos,
Universidad de Granada, Granada, Spain
\end{center}

\hfill

\centerline{\bf Abstract}

The determination of $|V_{us}|$ from kaon semileptonic decays requires the
value of the form factor $f_+(q^2=0)$ which can be calculated precisely on
the lattice. 
We provide the one-loop partially quenched chiral perturbation theory
expressions both with and without including the effects of staggered quarks
for all form factors at finite volume and with partially twisted
boundary conditions for both the vector current and scalar density
matrix elements at all $q^2$.
We point out that at finite volume there
are more form factors than just $f_+$ and $f_-$ for the vector current
matrix element but that the Ward identity is fully satisfied.
The size of the finite-volume corrections at present lattice sizes is small.  
This will help improve the lattice determination of $f_+(q^2=0)$ 
since the finite-volume error is the dominant error source for some calculations. 
The size of the finite-volume corrections may be estimated on a single lattice 
ensemble by comparing results for various twist choices.

\end{titlepage}

\section{Introduction}

The elements of the Cabibbo-Kobayashi-Maskawa (CKM) quark-mixing matrix are
fundamental parameters of the Standard Model (SM). The matrix is unitary in
the SM. Any deviation from unitarity would be a clear signal for new physics.
The first row, containing $V_{ud}$, $V_{us}$ and $V_{ub}$, is the one best
determined by experiment. For testing the unitarity relation
$|V_{ud}|^2+|V_{us}|^2+|V_{ub}|^2=1$, the precision on $|V_{ud}|$ and $|V_{us}|$ are
comparable \cite{Bazavov:2013maa}, while $|V_{ub}|$ is negligible at the current
level of precision. The determination of $|V_{us}|$ from semileptonic
kaon decays requires $f_+(q^2)$, the vector form factor of the $K$ to $\pi$ 
transition (see e.g. Ref.~\cite{Cirigliano:2011ny}). The ratio $f_+(q^2)/f_+(0)$ can 
be extracted from experiment, whereas theoretical input is needed for the absolute 
normalization given by the vector form factor at zero momentum transfer, $f_+(0)$.

The vector form factor is defined via
\begin{align}
  \label{eq:infVol}
 \left<\pi(p_\pi)|V_\mu|K(p_K)\right> = (p_K+p_\pi)_\mu f_+(q^2) + (p_K-p_\pi)_\mu f_-(q^2)
\end{align}
where $q=p_K-p_\pi$ and $V_\mu = \bar s\gamma_\mu q$, with $q$ the relevant
light quark.  The most precise way of calculating
$f_+(0)$ at present is with numerical
lattice QCD \cite{Bazavov:2013maa,Kaneko:2012cta,Boyle:2013gsa,KtopiLat2013,Boyle:2015hfa,Carrasco:2016kpy,Aoki:2016frl,Gamiz:2016bpm}. In lattice QCD calculations, as well as
experimentally, it is helpful to introduce the scalar form factor
\begin{align}
  f_0(q^2) = f_+(q^2)+f_-(q^2)\frac{q^2}{m_K^2-m_\pi^2}\,,
\end{align}
which satisfies
\begin{align}
  f_0(0) = f_+(0).
\end{align}
The form factors
$f_+$ and $f_0$ are less correlated than $f_+$ and $f_-$ and therefore easier
to disentangle experimentally. From a lattice perspective the scalar
form factor can be calculated using an insertion of a scalar current instead
of a vector current. Using a chiral Ward identity at zero momentum transfer
we have
\begin{align}
\label{WIrho}
  f_+(0) = f_0(0) = \frac{m_s-m_q}{m_K^2-m_\pi^2}\left<\pi(p_\pi)|S|K(p_K)\right>
\end{align}
where $S=\bar s q$. The scalar form factor is often easier to calculate
on the lattice.
Moreover, in the staggered formulation the local vector current  is not a
taste singlet and the added complications typically lead to larger
statistical errors \cite{Bazavov:2012cd,Na:2009au,Koponen:2012di}. 

Chiral perturbation theory (\chpt), with its various extensions to 
include discretization, finite-volume, and boundary-condition effects, 
plays an important role in handling the systematic errors of a lattice 
computation of $f_+(0)$. 
In this paper we calculate the finite-volume corrections to the
vector and scalar form factors in rooted staggered partially quenched {\chpt}
as well as in continuum \chpt. We also consider the effect of having 
twisted boundary conditions, possibly different for valence and sea quarks, 
on the finite-volume corrections. 
The infinite volume rooted staggered case is included in the calculation in the 
sense that results for that case can be obtained from our expressions by 
replacing finite-volume integrals by infinite-volume integrals, some of 
which are zero.

In a previous paper \cite{Bernard:2013eya} some of us developed a mixed action
formalism for staggered quarks. However, since the MILC collaboration has
moved to using only the highly-improved-staggered-quarks (HISQ) action, 
no such results are presented here. 
Some previous work on vector form factors in finite volume appears in 
Refs.~\cite{Ghorbani,Bijnens:2014yya,Jiang:2006gna}.

We point out that at finite volume there are more form factors than the
usual $f_+$ and $f_-$, which means that care has to be taken
while analysing Ward identities. In particular Eq.~(\ref{WIrho})
has corrections at finite volume and twisted boundary conditions.
We also point out that the finite-volume corrections
can be checked using only a single lattice ensemble by varying the
twisted boundary conditions.

We have implemented the resulting expressions numerically and they
will be made available in the CHIRON package \cite{CHIRON}. We have applied
the numerical programs to a set of ensembles from the MILC collaboration's 
HISQ ensembles \cite{Bazavov:2012xda} to show expected sizes of the corrections.
The main conclusions are that the finite-volume corrections are small
for present lattices. 

This paper is best read together with Ref.~\cite{Bernard:2013eya}
and is organized as follows: Section \ref{sec:chpt} establishes our conventions
and introduces the various versions of {\chpt} that we use. 
Section \ref{sec:Klrparam}
introduces our notation for the kaon semileptonic ($K_{l3}$) decays
and specifies the corrections to Eq.~(\ref{WIrho}) at finite volume.
Our analytical expressions for the $K_{l3}$ form factors are presented in
section \ref{sec:analytical}
 and some numerical examples are given in section \ref{sec:numerical}. 
Finally, section \ref{sec:conclusions} contains our conclusions. 
The integral notation used in our results, some integral identities, and 
additional results for meson masses and for form factors in the isospin limit 
can be found in the appendices. 
A preliminary version of this manuscript appeared in the PhD thesis
of Johan Relefors \cite{thesisjohan}.

\section{{\chpt} and lattice extensions}
\label{sec:chpt}

This section establishes our conventions and describes the lattice effects that we take into account. We start by introducing SU(3) {\chpt} in the continuum and then give the additional features needed for partially quenched {\chpt}, rooted staggered {\chpt} and twisted boundary conditions.
The conventions used are the same as in Ref.~\cite{Bernard:2013eya}. We work exclusively in Euclidean space.

Continuum infinite volume {\chpt} describes low energy QCD as an expansion in momenta and masses \cite{Weinberg,GL1,GL2}. It was first used in Ref.~\cite{GL3} to study meson form factors. The same Lagrangian can also be used in finite volume \cite{GL4}. In this paper we perform calculations to next-to-leading order (NLO), or $\mathcal{O}(p^4)$. The Lagrangian up to NLO is
\begin{align}
  \Lag = \Lag_2+\Lag_4
\end{align}
where $\Lag_{2n}$ is the $\mathcal{O}(p^{2n})$ Lagrangian. 

The effective degrees of freedom in the SU(3) case are the $\pi$, $K$, and $\eta$ mesons. For the fields we use the exponential representation
\begin{align}
  \label{eq:sigma}
  \Sigma = \exp\left(i\frac{2\phi}{f}\right),\, \text{with} \, 
  \phi = 
  \left(\begin{matrix}
      U & \pi^+ & K^+ \\
      \pi^- & D & K^0 \\
      K^- & \bar{K}^0 & S
    \end{matrix}\right),
\end{align}
where $f$ is the pion decay constant at LO and $U$, $D$ and $S$ are flavor neutral mesons with up, down and strange flavor respectively. 

The lowest order {\chpt} Lagrangian with external sources \cite{GL1,GL2} is given by
\begin{align}
  \label{eq:lag2}
  \Lag_2 = 
  \frac{f^2}{8}\tr\left(D_\mu\Sigma D_\mu\Sigma^\dagger\right) 
  - \frac{1}{4}\mu f^2\tr\left(\chi^\dagger\Sigma + \chi\Sigma^\dagger\right)
  + \frac{m_0^2}{6}\tr(\phi)^2
\end{align}
where $\mu$ is a low energy constant (LEC) and $\chi = s + ip$ contains scalar and pseudo scalar external fields. The covariant derivative is given by
\begin{align}
  D_\mu\Sigma = \partial_\mu \Sigma - i l_\mu\Sigma + i \Sigma r_\mu.
\end{align}
In order to include quark masses we let $s \rightarrow s + \text{diag}(m_u,m_d,m_s)$. The last term in $\Lag_2$ is essentially an $\eta^\prime$ mass term allowed by the anomaly. The mass should be taken to infinity in order to integrate out the $\eta^\prime$. This may be postponed until the final stage of the calculation \cite{Sharpe:2001fh}.  Postponing the limit is useful when discussing lattice effects since there is then a one-to-one relation between indices on $\phi$ and the quark content of the mesons \cite{Sharpe:1992ft}. When $m_{\eta^\prime}=m_0\rightarrow\infty$ the trace of $\phi$ decouples leaving $\pi_0$ and $\eta$ in the diagonal elements of $\phi$ and the correspondence is lost as standard {\chpt} is recovered. An expression for $\Lag_4$ can be found in Ref.~\cite{GL1}.

\subsection{Partially quenched \chpt}
\label{sec:PQ}
In partially quenched QCD the masses of the valence quarks differ from the masses of the sea quarks. In {\chpt} this can be incorporated using the observation that the indices on the meson matrix $\phi$ are quark indices before taking the limit $m_{\eta^\prime}\rightarrow \infty$. In a given diagram the indices that are determined by the external meson indices correspond to valence quarks, and we refer to these indices as valence indices. Indices that are summed over in a given diagram correspond to sea quarks, and we refer to these as sea indices. In this way there are sea-sea, sea-valence, valence-sea and valence-valence mesons.

From a technical point of view the partial quenching can be incorporated in {\chpt} using either the supersymmetric method \cite{Bernard:1993sv}, the replica method \cite{Damgaard:2000gh}, or using quark flow~\cite{Sharpe:1992ft}. The three methods give the same results in the partially quenched case (at least to one loop), but for the rooting of staggered quarks only the replica method or quark flow are applicable. As explained below, we find the quark-flow method more convenient. 
For this reason we have used the quark-flow method in our calculations.

From a calculational point of view one difference between standard {\chpt} and partially quenched {\chpt} is that the flavor neutral propagators have a more complicated structure. The flavor charged propagators have the standard form
\begin{align}
  G^C_{ef} = \frac{1}{p^2+m_{ef}^2}
\end{align}
where $e$ and $f$ indicate the flavor content of the meson. The flavor neutral propagators on the other hand have the form 
\begin{align}
  G^N_{EF} = G_{0,EF} + \mathcal{D}_{EF}
\end{align}
where 
\begin{align}
  \label{eq:PQprop}
  G_{0,EF} &= \frac{\delta_{EF}}{p^2+m_E^2}\textrm{ and}\\
  \mathcal{D}_{EF} &= - \frac{m_0^2}{3(p^2+m_E^2)(p^2+m_F^2)}\frac{(p^2+m_{U}^2)(p^2+m_{D}^2)(p^2+m_{S}^2)}{(p^2+m_\pi^2)(p^2+m_\eta^2)(p^2+m_{\eta^\prime}^2)}.\nonumber
\end{align}

In Eq.~(\ref{eq:PQprop}) $m_{U,D,S}$ are the masses of the neutral sea mesons with quark content 
$u\bar u$, $d\bar d$ and $s\bar s$, respectively, and $m_{\pi,\eta,\eta^\prime}$ are the masses of the $\pi,\eta,\eta^\prime$ sea mesons. $E$ and $F$ label flavor-neutral mesons (sea or valence). Note that $G^N_{EF}$ takes the form of a standard propagator plus a term due to the vertex proportional to $m_0^2$ of the type $\phi_{E}\phi_{F}$. We will refer to this type of vertex as a hairpin vertex. Letting $m_{\eta^\prime} = m_0 \rightarrow \infty$ gives \cite{Sharpe:2001fh}
\begin{align}
  \mathcal{D}_{EF} &= - \frac{1}{3(p^2+m_E^2)(p^2+m_F^2)}\frac{(p^2+m_{U}^2)(p^2+m_{D}^2)(p^2+m_{S}^2)}{(p^2+m_\pi^2)(p^2+m_\eta^2)}.
\end{align}

\subsection{Rooted staggered \chpt}
\label{sec:staggered}
We now introduce staggered quarks and rooting in {\chpt}. In the staggered formulation of lattice QCD each quark is fourfold degenerate. In lattice simulations this is compensated for by taking the fourth root of the quark determinant, the so called fourth root trick. A consequence of the fourfold degeneracy is that the number of mesons is increased 16 fold, giving 16 tastes for each flavor. In staggered {\chpt} the degeneracy is compensated for by dividing each sum over sea quarks by four, mimicking the fourth-root trick. This is the reason why having a direct correspondence between the indices of $\phi$ and the quark content of the corresponding meson is so useful when dealing with staggered quarks. Also, note that in the replica method any summed over flavor index is a sea index so that each sum should simply be divided by four.

In order to accommodate the 16 fold increase in the number of mesons in {\chpt} we use the representation
\begin{align}
  \label{eq:staggSigma}
  \Sigma = \exp\left(i\frac{\phi}{f}\right),\quad \text{with}\quad
  \phi = 
  \left(\begin{matrix}
      U & \pi^+ & K^+ &\hdots\\
      \pi^- & D & K^0 &\hdots\\
      K^- & \bar{K}^0 & S & \hdots\\
      \vdots &\vdots&\vdots&\ddots    
    \end{matrix}\right),
\end{align}
where the extra space in the matrix $\phi$ can be used to accommodate partial quenching \cite{Bernard:1993sv,Damgaard:2000gh}. Each entry in $\phi$ is a $4\times 4$ matrix written as 
\begin{align}
  \pi^a \equiv \sum_{\Xi = 1}^{16} \pi^a_\Xi T_\Xi\,,
\quad \text{where}\quad
 T_\Xi\in\left\{\xi_5,i\xi_{\mu 5},i\xi_{\mu\nu} (\mu>\nu),\xi_\mu,I \right\}
\end{align}
are the taste generators, here taken as the Euclidean gamma matrices $\xi_\mu$, with $\xi_{\mu\nu} = \xi_\mu\xi_\nu$, $\xi_{\mu 5} \equiv \xi_\mu\xi_5$ and $\xi_I\equiv I$ is the $4\times 4$ identity matrix. These generate $U(4)$ which is the coset space of a single flavor staggered theory where the trace is not decoupled. The tastes of mesons will also be referred to as P,A,T,V and I. As long as no discretization effects are taken into account all tastes with the same flavor have degenerate masses, this degeneracy is broken by discretization effects.

When including effects from the lattice spacing $a$, we treat $p^2$, $m_q$ and $a^2$ as the same order in our power counting. $\Lag_2$ will then contain corrections of $\mathcal{O}(a^2)$. Although such effects break  the 16 fold degeneracy in the meson spectrum, it turns out \cite{Lee:1999zxa} that at this order in the power counting there is still an $SO(4)$ symmetry, sometimes referred to as the residual taste symmetry. Breaking of $SU(4)$ to $SO(4)$ lifts the degeneracy of mass between the tastes P,A,T,V and I, giving five different masses for each meson flavor.

Using the conventions in Ref.~\cite{Bernard:2013eya}, the Lee-Sharpe Lagrangian \cite{Lee:1999zxa} generalized to multiple flavors \cite{Aubin:2003mg} is written as
\begin{align}
  \mathcal{L} = 
  \frac{f^2}{8}\tr\left(D_\mu\Sigma D_\mu\Sigma^\dagger\right) 
  - \frac{1}{4}\mu f^2\tr\left(\chi^\dagger\Sigma + \chi\Sigma^\dagger\right)
  +\frac{m_0^2}{24}\left(\tr\left(\Phi^2\right)\right)
  +a^2\mathcal{V}\,,
\end{align}
where $\mathcal{V}$ is the taste violating potential found in Ref.~\cite{Aubin:2003mg}. The $m_0^2$ term is the contribution to the singlet-taste and singlet-flavor meson, $\eta^\prime_I\propto \tr(\phi)$, which is the only mass term allowed by the anomaly. As in the continuum partially quenched case, the limit $m_0\rightarrow\infty$ can be taken at the end of the calculation in order to keep a correspondence between the indices of $\phi$ and the quark content of the mesons. 

The flavor neutral propagators are again more complicated than in standard {\chpt}. In the staggered theory the $m_0^2$ terms generate hairpin vertices for the singlet-taste flavor-neutral mesons. There are also hairpin vertices for the axial and vector taste flavor neutral mesons coming from double trace terms in the staggered Lagrangian. The neutral propagators for taste $\Xi$ are in this case given by
\begin{align}
  G^N_{EF,\Xi} = G_{0,EF,\Xi} + \mathcal{D}_{EF}^\Xi
\end{align}
where 
\begin{align}
  \label{eq:staggProp}
  G_{0,EF,\Xi} &= \frac{\delta_{EF}}{p^2+m_{A,\Xi}^2}\textrm{ and}\\
  \mathcal{D}^{\Xi}_{EF} &= -a^2 \delta_{\Xi}
  \frac
  {(p^2+m_{U,\Xi}^2)(p^2+m_{D,\Xi}^2)(p^2+m_{S,\Xi}^2)}
  {(p^2+m_{E,\Xi}^2)(p^2+m_{F,\Xi}^2)(p^2+m_{\pi^0,\Xi}^2)(p^2+m_{\eta,\Xi}^2)(p^2+m_{\eta^\prime,\Xi}^2)}.\nonumber
\end{align}
In Eq.~(\ref{eq:staggProp}) $\delta_{\Xi}$ are the couplings for the hairpin vertices, for tastes $\Xi=V,A,I$ respectively. In the limit $m_0\rightarrow\infty$ the singlet-taste disconnected flavor-neutral propagator simplifies to
\begin{align}
  \mathcal{D}^{I}_{EF} = -\frac{4}{3}
  \frac
  {(p^2+m_{U,I}^2)(p^2+m_{D,I}^2)(p^2+m_{S,I}^2)}
  {(p^2+m_{A,I}^2)(p^2+m_{B,I}^2)(p^2+m_{\pi^0,I}^2)(p^2+m_{\eta,I}^2)}.
\end{align}
The other tastes have no hairpin vertices and hence $\mathcal{D}^{T,P}=0$.

\subsection{Twisted boundary conditions}
Twisted boundary conditions \cite{Bedaque} in one dimension are defined by
\begin{align}
  \label{eq:defTwist}
  \psi(x+L) = \exp(i\theta)\psi(x)
\end{align}
where $L$ is the length of the dimension and $\theta$ is the twist angle. With twisted boundary conditions momenta are quantized as
\begin{align}
  p = \frac{2\pi}{L}n +\frac{\theta}{L},\quad n\in\mathrm{Z}.
\end{align}
The twist angle can be chosen arbitrarily, so the momentum of the field $\psi$ can be continuously varied. In the case $\theta=0$, periodic boundary conditions are recovered. The twist of the anti-particle follows from complex conjugation of (\ref{eq:defTwist}); momenta are shifted in the opposite direction. 

Twist angles can be chosen independently in each spatial direction for each flavor and also independently for sea and valence quarks. For each quark $q$, either valence or sea, we define the twist angle, $\theta_i^q$, in direction $i$ via
\begin{align}
  q(x_i+L) = \exp(i\theta_i^q)q(x_i).
\end{align}
We collect the twist angles $\theta_i^q$ in a three vector $\vec{\theta}^q$ and in a four vector $\theta^q = (0,\vec \theta^q)$. The twist angle for an anti-quark is minus the twist angle for the corresponding quark.

The twist angles of the mesons follow from that of the quarks as \cite{twisted}
\begin{align}
  \phi_{\bar q^\prime q}(x_i+L) = \exp(i(\theta_i^q-\theta_i^{q^\prime}))\phi_{\bar q^\prime q}(x_i)
\end{align}
where $\phi_{\bar q^\prime q}$ is a meson with quark content $\bar q^\prime q$. It follows that flavor-diagonal mesons have zero twist angle and that charge-conjugate mesons have opposite twist. A particle with spatial momentum $\vec p$ has an anti-particle with spatial momentum $-\vec p$.

When computing loop integrals using twisted boundary conditions in a finite volume we have to make the replacement
\begin{align}
  \label{eq:twistInt}
  \int \frac{d^dk}{(2\pi)^d} \rightarrow \int_V\frac{d^dk}{(2\pi)^d} \equiv \int \frac{d^{d-3}k}{(2\pi)^{d-3}}\frac{1}{L^3}\sum_{\substack{\vec n\in \mathrm{Z}^3 \\\vec k=(2\pi\vec n+\vec\theta)/L}}
\end{align}
where we allow for dimensional regularization by using a total of $d$ dimensions. Note that the twisted boundary conditions lead to
\begin{align}
  \int_V\frac{d^dk}{(2\pi)^d}\frac{k_\mu}{k^2+m^2} \neq 0
\end{align}
since the sum is not symmetric around zero. This leads to momentum-dependent masses and fewer constraints on form factors, which reflects the broken lattice symmetry. This also makes checking Ward identities more involved than in the usual case \cite{Bijnens:2014yya}.

\section{Parameterization of kaon semileptonic decays at finite volume}
\label{sec:Klrparam}

In this section we present our calculation of the finite-volume corrections for the hadronic matrix element in $K_{l3}$ decays. Although we use $K^0 \rightarrow \pi^- l^+\nu$ as an example, our calculations can be used for any  $K \rightarrow \pi l\nu$ decay. At the quark level, the decay $K^0 \rightarrow \pi^- l^+\nu$ is due to the vector current $\bar s \gamma_\mu u$. In order to keep the discussion general we follow Ref.~\cite{Bernard:2013eya} and define $\bar y$ and $\bar x$ to be the valence anti-quarks corresponding to $\bar s$ and $\bar u$ respectively. We also define $x^\prime$ to be the spectator valence quark corresponding to the $d$ quark. The decay is then that of an $x^\prime\bar y$ to an $x^\prime\bar x$ pseudo scalar through the vector current $\bar y \gamma_\mu x$. We also introduce the notation $X$, $X^\prime$ and $Y$ for the valance pseudo scalar mesons $x\bar x$, $x^\prime \bar x^\prime$ and $y \bar y$.

We parameterize the matrix element of the weak current between a kaon and a pion in finite volume as
\begin{align}
  \left<\pi(p_\pi)|V_\mu^{xy}|K(p_K)\right>_V = f^{xy}_+(q)(p_K+p_\pi)_\mu + f^{xy}_-(q)(p_K-p_\pi)_\mu + h^{xy}_\mu(q),
\end{align}
where $q= (p_K-p_\pi)$ and $V_\mu^{xy}$ is the appropriate flavor-changing vector current. In the various versions of {\chpt} presented above $V_\mu^{xy}$ share the same form given by
\begin{align}
  V_\mu^{xy} = \frac{if^2}{4}\tr_t\left[\partial_\mu\Sigma \Sigma^\dagger - \Sigma^\dagger\partial_\mu\Sigma\right]_{xy},
\end{align}
where the content of $\Sigma$ will differ in the different versions, and $\tr_t$ is a trace over taste only (which simply gives one in the non-staggered theory). Our conventions are such that $f_+ = 1$ at leading order in {\chpt}. For zero twist angle the restored cubic symmetry means that only the first two terms are needed so that $h_\mu=0$ in this case. For non-zero twist angle $h_\mu\neq 0$. Note that the split between different form factors is not unique in this case. For example, changing routings in a diagram will shift terms between $f_-$ and $h_\mu$. Also, the form-factors depend on the individual components of $q$ through the twist angles which enter the integrals, see Ref.~\cite{Bijnens:2014yya}.
Nevertheless, although the split is in some sense artificial when twisted boundary conditions are imposed, it is useful in order to relate to the infinite volume limit where there are well defined form factors depending only on $q^2$; see Eq.~(\ref{eq:infVol}).

In practice it is advantageous to study the scalar form factor on the lattice and then relate the result to the vector form factor \cite{Bazavov:2012cd,Na:2010uf}. In {\chpt} the scalar current is
\begin{align}
  S_{xy} = -\frac{f^2\mu}{4}\tr_t\left(\Sigma+\Sigma^\dagger\right)_{xy}.
\end{align}
We parameterize the matrix element between a kaon and a pion as
\begin{align}
  \left<\pi(p_\pi)|S_{xy}|K(p_K)\right>_V = \frac{\rho_{xy}(q)}{m_y-m_x}.
\end{align}
With these definitions the Ward-Takahashi identity relating the hadronic matrix elements leads to the following relation between the relevant form factors
\begin{align}
  \label{eq:ward}
  (p_K^2-p_\pi^2)f_+^{xy}+q^2f_-^{xy}+q_\mu h_\mu^{xy} = -\rho^{xy}.
\end{align}
Note that $p_{K/\pi}^2$ must contain the full loop contribution, to the order at which the Ward identity is being checked, since $f_+=1$ at leading order. In all results presented below we have checked that this Ward identity holds.

Finally, setting $q^2=0$, which is important for $|V_{us}|$, we have the relation
\begin{align}
  f_+^{xy}(q^2=0) = \left.\frac{-\rho^{xy}-q_\mu h_\mu^{xy}}{(p_K^2-p_\pi^2)}\right|_{q^2=0}
\end{align}
where $h_\mu^{xy}$ vanishes in the infinite volume limit, allowing for a determination of the vector form factor from the scalar form factor. In lattice calculations the term proportional to $q_\mu h_\mu$ is often dropped \cite{Bazavov:2013maa,Kaneko:2012cta,Boyle:2013gsa,Boyle:2015hfa,Carrasco:2016kpy,Aoki:2016frl,Gamiz:2016bpm,Bazavov:2012cd,Kaneko:2016mha}. The left hand side of the equation is then not $f_+$ but a quantity which goes to $f_+$ in the infinite volume limit.

\section{Finite-volume corrections to $f_+$, $f_-$, $h_\mu$ and $\rho$}
\label{sec:analytical}

In this section we present finite-volume corrections to the hadronic matrix
elements needed for $K_{l3}$ decays at NLO in {\chpt}. We present rooted
staggered partially quenched ChPT (rSPQChPT) expressions for the partially
twisted case (twisted boundary conditions different in the valence and sea
sectors), as well as the corresponding continuum limit
(PQChPT with partially twisted boundary conditions). The continuum limit can
be derived from the staggered results, but we present both for clarity. 
The finite-volume corrections can be used to derive the infinite-volume
expressions. To do this one should replace every finite-volume integral by its
infinite-volume counterpart. The expressions are presented using the 
$\mathcal D$ notation of Ref.~\cite{Bernard:2013eya}, which keep the diagonal
propagators intact, see Appendix \ref{sec:integrals}. This is to keep
the expressions of manageable length. 

Taking the full QCD infinite volume and isospin limits of the PQ result
produces a slightly different expression from the NLO results
in Ref.~\cite{Bijnens:2003uy}. The difference is of $\mathcal{O}(p^6)$. There is,
however, no conflict in using our result for the finite-volume corrections with 
the infinite-volume NLO+NNLO calculation of Ref.~\cite{Bijnens:2003uy} since 
there is no overlap between the finite- and infinite-volume results.

 Below, we give the finite-volume corrections to hadronic matrix elements of both vector and scalar currents. For a given quantity, $X$, the finite-volume correction, $\Delta^VX$, is defined as
\begin{align}
  \Delta^VX = X^V - X^\infty
\end{align}
where $X^V$ is $X$ calculated in finite volume and $X^\infty$ is $X$ calculated in infinite volume. We envision computing $X^V$ on the lattice and subtracting $\Delta^V X$ to correct for finite-volume effects, thereby obtaining $X^\infty$, the quantity of interest. 
The case of $h_\mu^V$ is special in that the corresponding infinite-volume expression is zero.

The finite-volume expressions depend on the volume through the integrals $A^V$, $B^V$, etc. These integrals also depend on the masses and twist angles of both valence and sea quarks. In staggered {\chpt} there are additional low energy constants which enter through the relation between meson masses and quark masses and through hairpin couplings for the diagonal vector and axial propagators.

In the staggered case, we take the external mesons to be taste pseudoscalars (taste $\xi_5$), as in Ref.~\cite{Bernard:2013eya}. The quantity $c_\Xi$, defined as
\begin{align}
  c_\Xi = \frac{1}{4}\textrm{Tr}(\xi^5\xi^\Xi\xi^5\xi^\Xi)\,,
\end{align}
then appears in the rooted staggered expressions. In addition to $q = p_K-p_\pi$, we use the momentum variable
\begin{align}
  p_{12} = p_K+p_\pi.\nonumber
\end{align}

Some complementary results have been moved to the Appendix. 
Appendix \ref{sec:finiteVolumeMasses} presents results for the finite-volume correction to the masses in the partially-twisted partially-quenched and partially-twisted partially-quenched rooted staggered cases. These are needed to check the Ward identity in Eq.~(\ref{eq:ward}). 
In Appendix \ref{sec:isospinKl3} we give expressions for the partially-twisted and 
fully-twisted $K^0 \rightarrow \pi^-$ form factors in the isospin
limit, in which most of the current lattice calculations are performed. 

\subsection{Continuum Partially-Quenched Partially-Twisted \chpt}

 Here we present results for the finite-volume corrections to the $K_{l3}$ form factors, calculated using PQ{\chpt} at $\mathcal O(p^4)$, when the inserted current is a vector current (Sec. \ref{sec:PQvec}) and a scalar current (Sec. \ref{sec:PQscal}). 

\subsubsection{Finite-volume corrections for the vector form factors}
\label{sec:PQvec}

\begin{align}
  \Delta^V f_+^{xy} = -\frac{1}{2f^2}\left(\vphantom{\frac{1}{2}}\right.
  &\sum_{\mathcal{S}}\left[-A^{V}_{}(m^2_{y\mathcal{S}})
    -A^{V}_{}(m^2_{x\mathcal{S}})
    +4B^{V}_{22}(m^2_{x\mathcal{S}},m^2_{\mathcal{S}y},q)\right]\\
  +&4\left[B^{V}_{22}(m^2_{xy},\mathcal{D}_{{Y}{Y}},q)
  -2B^{V}_{22}(m^2_{xy},\mathcal{D}_{{Y}{X}},q)+B^{V}_{22}(m^2_{xy},\mathcal{D}_{{X}{X}},q) \right]\nonumber\\
%  % 
%  \right) \nonumber\\
  % 
  -&A^{V}_{}(\mathcal{D}_{{Y}{Y}})
  +2A^{V}_{}(\mathcal{D}_{{Y}{X}})
  -A^{V}_{}(\mathcal{D}_{{X}{X}})
  \left.\vphantom{\frac{1}{2}}\right)\nonumber
\end{align}
\begin{align}
  \Delta^V f_-^{xy}= -\frac{1}{2f^2}\left(\vphantom{\frac{1}{2}}\right.
  %%%%%%%%%%%%%%%%%%%%%%%%%%%%
  %%%%%%%%%%%%%%%%%%%%%%%%%%%%
  &\sum_{\mathcal{S}}\left[4\left(m^2_{x^\prime y}-m^2_{x^\prime x}\right)\left\{B^{V}_{21}(m^2_{x\mathcal{S}},m^2_{\mathcal{S}y},q)\right.\right.\\
  &\hphantom{\sum_{\mathcal{S}}(4(m^2_{x^\prime y}-m^2_{x^\prime x})}\left.-B^{V}_{1}(m^2_{x\mathcal{S}},m^2_{\mathcal{S}y},q)\right\}\nonumber\\
  %%%%%%%%%%%%%%%%%%%%%%%%%%%%
  %%%%%%%%%%%%%%%%%%%%%%%%%%%%
  &\left.\hphantom{\sum_{\mathcal{S}}(}+2q_{\mu}B^{V}_{2\mu}(m^2_{x\mathcal{S}},m^2_{\mathcal{S}y},q) +2p_{12\mu}B^{V}_{2\mu}(m^2_{x\mathcal{S}},m^2_{\mathcal{S}y},q)\right] \nonumber\\
  %%%%%%%%%%%%%%%%%%%%%%%%%%%%
  %%%%%%%%%%%%%%%%%%%%%%%%%%%%
  +&4\left(m^2_{x^\prime y}-m^2_{x^\prime x}\right)\left[
    B^{V}_{21}(m^2_{xy},\mathcal{D}_{{Y}{Y}},q)
    -2B^{V}_{21}(m^2_{xy},\mathcal{D}_{{Y}{X}},q)\right.\nonumber\\
  &\hphantom{4\left(m^2_{x^\prime y}-m^2_{x^\prime x}\right)}\left.
    +B^{V}_{21}(m^2_{xy},\mathcal{D}_{{X}{X}},q)
      \right]  \nonumber\\
  %%%%%%%%%%%%%%%%%%%%%%%%%%%%
  %%%%%%%%%%%%%%%%%%%%%%%%%%%%
  +&4B^{V}_{1}(m^2_{xy},\mathcal{D}_{{X}^\prime{Y}},q)\left(
    -2m^2_{x^\prime x^\prime}
    +3m^2_{x^\prime y}
    +m^2_{x^\prime x}
  \right) \nonumber\\
  %%%%%%%%%%%%%%%%%%%%%%%%%%%%
  %%%%%%%%%%%%%%%%%%%%%%%%%%%%
  +&4B^{V}_{1}(m^2_{xy},\mathcal{D}_{{X}^\prime{X}},q)\left(
    2m^2_{x^\prime x^\prime}
    -m^2_{x^\prime y}
    -3m^2_{x^\prime x}
  \right) \nonumber\\
  %%%%%%%%%%%%%%%%%%%%%%%%%%%%
  %%%%%%%%%%%%%%%%%%%%%%%%%%%%
  -&4\left(m^2_{x^\prime y}-m^2_{x^\prime x}\right)\left[
    B^{V}_{1}(m^2_{xy},\mathcal{D}_{{Y}{Y}},q)
    +B^{V}_{1}(m^2_{xy},\mathcal{D}_{{X}{X}},q)\right]\nonumber\\
  %%%%%%%%%%%%%%%%%%%%%%%%%%%%
  %%%%%%%%%%%%%%%%%%%%%%%%%%%%
  -&4q_{\mu}B^{V}_{2\mu}(m^2_{xy},\mathcal{D}_{{X}^\prime{Y}},q)
  +4q_{\mu}B^{V}_{2\mu}(m^2_{xy},\mathcal{D}_{{X}^\prime{X}},q) \nonumber\\
  +&2q_{\mu}B^{V}_{2\mu}(m^2_{xy},\mathcal{D}_{{Y}{Y}},q)
  -2q_{\mu}B^{V}_{2\mu}(m^2_{xy},\mathcal{D}_{{X}{X}},q)\nonumber\\
  % 
  %%%%%%%%%%%%%%%%%%%%%%%%%%%%
  %%%%%%%%%%%%%%%%%%%%%%%%%%%%
  +&2p_{12\mu}B^{V}_{2\mu}(m^2_{xy},\mathcal{D}_{{Y}{Y}},q)
  -4p_{12\mu}B^{V}_{2\mu}(m^2_{xy},\mathcal{D}_{{Y}{X}},q)\nonumber\\
  +&2p_{12\mu}B^{V}_{2\mu}(m^2_{xy},\mathcal{D}_{{X}{X}},q)
  \nonumber\\
  %%%%%%%%%%%%%%%%%%%%%%%%%%%%
  %%%%%%%%%%%%%%%%%%%%%%%%%%%%
  +&4B^{V}_{}(m^2_{xy},\mathcal{D}_{{X}^\prime{Y}},q)\left(
    m^2_{x^\prime x^\prime}
    -m^2_{x^\prime y}
    -m^2_{x^\prime x}
  \right) \nonumber\\
  %%%%%%%%%%%%%%%%%%%%%%%%%%%%
  %%%%%%%%%%%%%%%%%%%%%%%%%%%%
  +&4B^{V}_{}(m^2_{xy},\mathcal{D}_{{X}^\prime{X}},q)\left(
    -m^2_{x^\prime x^\prime}
    +m^2_{x^\prime y}
    +m^2_{x^\prime x}
  \right) 
  \left.\vphantom{\frac{1}{2}}\right)\nonumber
\end{align}
\begin{align}
  \Delta^V h_\mu^{xy}= -\frac{1}{2f^2}\left(\vphantom{\frac{1}{2}}\right.
  &\sum_{\mathcal{S}}\left[-4p_{12\nu}B^{V}_{23\mu\nu}(m^2_{x\mathcal{S}},m^2_{\mathcal{S}y},q)\right. \\
  &\hphantom{\sum_{\mathcal{S}}(}+2B^{V}_{2\mu}(m^2_{x\mathcal{S}},m^2_{\mathcal{S}y},q)\left(
    -q^2
    -m^2_{x^\prime y}
    +m^2_{x^\prime x}
  \right) \nonumber\\
  &\hphantom{\sum_{\mathcal{S}}(}-4A^{V}_{\mu}(m^2_{x^\prime \mathcal{S}})
  +2A^{V}_{\mu}(m^2_{y\mathcal{S}})
  +2A^{V}_{\mu}(m^2_{x\mathcal{S}})\left.\vphantom{B^{V}_{23\mu\nu}}\right]\nonumber\\
  % 
  %%%%%%%%%%%%%%%%%%%%%%%%%%%%
  %%%%%%%%%%%%%%%%%%%%%%%%%%%%
  -&4p_{12\nu}\left[B^{V}_{23\mu\nu}(m^2_{xy},\mathcal{D}_{{Y}{Y}},q)+B^{V}_{23\mu\nu}(m^2_{xy},\mathcal{D}_{{X}{X}},q)\right] \nonumber\\
  +&8p_{12\nu}B^{V}_{23\mu\nu}(m^2_{xy},\mathcal{D}_{{Y}{X}},q) \nonumber\\
%  % 
%  -4p_{12\nu}& \nonumber\\
  %%%%%%%%%%%%%%%%%%%%%%%%%%%%
  %%%%%%%%%%%%%%%%%%%%%%%%%%%%
  +&4B^{V}_{2\mu}(m^2_{xy},\mathcal{D}_{{X}^\prime{Y}},q)\left(
    q^2
    -m^2_{x^\prime x^\prime}
    +2m^2_{x^\prime y}
    +m^2_{x^\prime x}
  \right) \nonumber\\
  %%%%%%%%%%%%%%%%%%%%%%%%%%%%
  %%%%%%%%%%%%%%%%%%%%%%%%%%%%
  +&4B^{V}_{2\mu}(m^2_{xy},\mathcal{D}_{{X}^\prime{X}},q)\left(
    -q^2
    +m^2_{x^\prime x^\prime}
    -m^2_{x^\prime y}
    -2m^2_{x^\prime x}
  \right) \nonumber\\
  %%%%%%%%%%%%%%%%%%%%%%%%%%%%
  %%%%%%%%%%%%%%%%%%%%%%%%%%%%
  +&2B^{V}_{2\mu}(m^2_{xy},\mathcal{D}_{{Y}{Y}},q)\left(
    -q^2
    -m^2_{x^\prime y}
    +m^2_{x^\prime x}
  \right) \nonumber\\
  %%%%%%%%%%%%%%%%%%%%%%%%%%%%
  %%%%%%%%%%%%%%%%%%%%%%%%%%%%
  +&2B^{V}_{2\mu}(m^2_{xy},\mathcal{D}_{{X}{X}},q)\left(
    q^2
    -m^2_{x^\prime y}
    +m^2_{x^\prime x}
  \right) 
  \left.\vphantom{\frac{1}{2}}\right)\nonumber
\end{align}

\subsubsection{Finite-volume corrections for the scalar form factor}
\label{sec:PQscal}
\begin{align}
  \frac{\Delta^V\rho_{xy}}{(m_K^2-m_\pi^2)} = -\frac{1}{2f^2}\left(\vphantom{\frac{1}{2}}\right.
  &\sum_{\mathcal{S}}\left[-2\left(m^2_{x^\prime y}-m^2_{x^\prime x}\right)B^V_{1}(m^2_{x\mathcal{S}},m^2_{\mathcal{S}y},q)\right. \\
  &\hphantom{\sum_{\mathcal{S}}(}+2p_{12\mu} B_{2}^{V\mu}(m^2_{x\mathcal{S}},m^2_{\mathcal{S}y},q)\nonumber\\
  &\left.\hphantom{\sum_{\mathcal{S}}(} +B^V(m^2_{x\mathcal{S}},m^2_{\mathcal{S}y},q)\left(
      q^2
      +m^2_{x^\prime y}
      -m^2_{x^\prime x}
    \right)\right] \nonumber\\
  %%%%%%%%%%%%%%%%%%%%%%%%%%%% 
  %%%%%%%%%%%%%%%%%%%%%%%%%%%% 
  -&2\left(m^2_{x^\prime y}-m^2_{x^\prime x}\right)\left[
    B^V_{1}(m^2_{xy},\mathcal{D}_{YY},q)\right.\nonumber\\
  &\hphantom{2\left(m^2_{x^\prime y}-m^2_{x^\prime x}\right)}
  -\left.B^V_{1}(m^2_{xy},\mathcal{D}_{XX},q)\right] \nonumber\\
  %%%%%%%%%%%%%%%%%%%%%%%%%%%%
  %%%%%%%%%%%%%%%%%%%%%%%%%%%%
  +&2p_{12\mu}B_{2}^{V\mu}(m^2_{xy},\mathcal{D}_{YY},q)
  -2p_{12\mu} B_{2}^{V\mu}(m^2_{xy},\mathcal{D}_{X X},q) \nonumber\\
  %%%%%%%%%%%%%%%%%%%%%%%%%%%%
  %%%%%%%%%%%%%%%%%%%%%%%%%%%%
  -& 2B^V(m^2_{xy},\mathcal{D}_{X^\prime Y},q)\left(
    q^2
    +m^2_{x^\prime y}
    +m^2_{x^\prime x}
  \right) \nonumber\\
  %%%%%%%%%%%%%%%%%%%%%%%%%%%%
  %%%%%%%%%%%%%%%%%%%%%%%%%%%%
  -& 2B^V(m^2_{xy},\mathcal{D}_{X^\prime X},q)\left(
    q^2
    +m^2_{x^\prime y}
    +m^2_{x^\prime x}
  \right) \nonumber\\
  %%%%%%%%%%%%%%%%%%%%%%%%%%%%
  %%%%%%%%%%%%%%%%%%%%%%%%%%%%
  +& B^V(m^2_{xy},\mathcal{D}_{YY},q)\left(
    q^2
    +m^2_{x^\prime y}
    -m^2_{x^\prime x}
  \right) \nonumber\\
  %%%%%%%%%%%%%%%%%%%%%%%%%%%%
  %%%%%%%%%%%%%%%%%%%%%%%%%%%%
  +&2 B^V(m^2_{xy},\mathcal{D}_{YX},q)
    q^2
   \nonumber\\
  +& B^V(m^2_{xy},\mathcal{D}_{XX},q)\left(
    q^2
    -m^2_{x^\prime y}
    +m^2_{x^\prime x}
  \right) \nonumber\\
  %%%%%%%%%%%%%%%%%%%%%%%%%%%% 
  %%%%%%%%%%%%%%%%%%%%%%%%%%%%
  -& 2A(\mathcal{D}_{X^\prime Y})
  - 2A(\mathcal{D}_{X^\prime X})
  \left.\vphantom{\frac{1}{2}}\right)\nonumber
\end{align}

\subsection{Partially-Quenched Partially-Twisted Rooted Staggered \chpt}

In this subsection, we give the finite-volume corrections to the $K_{l3}$ form factors, calculated using rSPQ{\chpt} at $\mathcal O(p^4)$, when the inserted current is a vector current (Sec. \ref{sec:rSPQvec}) and a scalar current (Sec. \ref{sec:rSPQscal}).

\subsubsection{Finite-volume corrections for the vector form factor}
\label{sec:rSPQvec}
\begin{align}
  \Delta^V f_+^{xy}= -\frac{1}{2f^2}\sum_\Xi \left(\vphantom{\frac{1}{2}}\right.
  &\frac{1}{16}\sum_{\mathcal{S}}\left[-A^{V}_{}(m^2_{y\mathcal{S},\Xi})-A^{V}_{}(m^2_{x\mathcal{S},\Xi})  +\frac{1}{4}B^{V}_{22}(m^2_{x\mathcal{S},\Xi},m^2_{\mathcal{S}y,\Xi},q)\right] \nonumber\\
  &+B^{V}_{22}(m^2_{xy,\Xi},\mathcal{D}^\Xi_{YY},q)
  -2B^{V}_{22}(m^2_{xy,\Xi},\mathcal{D}^\Xi_{YX},q)\\
  &+B^{V}_{22}(m^2_{xy,\Xi},\mathcal{D}^\Xi_{XX},q)\nonumber\\
  &-\frac{1}{4}\left[A^{V}_{}(\mathcal{D}^\Xi_{YY})-2A^{V}_{}(\mathcal{D}^\Xi_{YX})+A^{V}_{}(\mathcal{D}^\Xi_{XX})\right]  \left.\vphantom{\frac{1}{2}}\right)\nonumber
\end{align}
\begin{align}
  \Delta^V f_-^{xy}=-\frac{1}{2f^2}\sum_\Xi\left(\vphantom{\frac{1}{2}}\right.
  %%%%%%%%%%%%%%%%%%%%%%%%%%%%
  %%%%%%%%%%%%%%%%%%%%%%%%%%%%
  &\frac{1}{4}\sum_{\mathcal{S}}\left[\left(m^2_{x^{\prime}y,5}-m^2_{x^{\prime}x,5}\right)
    \left\{B^{V}_{21}(m^2_{x\mathcal{S},\Xi},m^2_{\mathcal{S}y,\Xi},q)\right.\right.\\
  &\hphantom{\sum_{\mathcal{S}}\left((m^2_{x^{\prime}y,5}-m^2_{x^{\prime}x,5}\right)(}\left.
  -\left.B^{V}_{1}(m^2_{x\mathcal{S},\Xi},m^2_{\mathcal{S}y,\Xi},q)\right\} \right.\nonumber\\
%  % 
  &\phantom{\sum_{\mathcal{S}}(}
  +\frac{q_{\mu}}{2}B^{V}_{2\mu}(m^2_{x\mathcal{S},\Xi},m^2_{\mathcal{S}y,\Xi},q)\nonumber\\
  &\phantom{\sum_{\mathcal{S}}(}
  +\frac{p_{12\mu}}{2}\left.B^{V}_{2\mu}(m^2_{x\mathcal{S},\Xi},m^2_{\mathcal{S}y,\Xi},q)
  \right]   \nonumber\\
  %%%%%%%%%%%%%%%%%%%%%%%%%%%%
  %%%%%%%%%%%%%%%%%%%%%%%%%%%%
  &+\left(m^2_{x^{\prime}y,5}-m^2_{x^{\prime}x,5}\right)\left[
    B^{V}_{21}(m^2_{xy,\Xi},\mathcal{D}^\Xi_{YY},q)
    -2B^{V}_{21}(m^2_{xy,\Xi},\mathcal{D}^\Xi_{YX},q) \right.\nonumber\\
  &\hphantom{+\left(m^2_{x^{\prime}y,5}-m^2_{x^{\prime}x,5}\right)}\left.
    +B^{V}_{21}(m^2_{xy,\Xi},\mathcal{D}^\Xi_{XX}
     \right] \nonumber\\
  %%%%%%%%%%%%%%%%%%%%%%%%%%%%
  %%%%%%%%%%%%%%%%%%%%%%%%%%%%
  +&B^{V}_{1}(m^2_{xy,\Xi},\mathcal{D}^\Xi_{X^{\prime}Y},q)\left(
    -2m^2_{x^{\prime}x^{\prime},5}
    +m^2_{x^{\prime}y,5}\left[2+c_\Xi\right]
    +m^2_{x^{\prime}x,5}c_\Xi
  \right) \nonumber\\
  %%%%%%%%%%%%%%%%%%%%%%%%%%%%
  %%%%%%%%%%%%%%%%%%%%%%%%%%%%
  +&B^{V}_{1}(m^2_{xy,\Xi},\mathcal{D}^\Xi_{X^{\prime}X},q)\left(
    2m^2_{x^{\prime}x^{\prime},5}
    -m^2_{x^{\prime}y,5}c_\Xi
    -m^2_{x^{\prime}x,5}\left[2+c_\Xi\right]
  \right) \nonumber\\
  %%%%%%%%%%%%%%%%%%%%%%%%%%%%
  %%%%%%%%%%%%%%%%%%%%%%%%%%%%
  -&\left(m^2_{x^{\prime}y,5}-m^2_{x^{\prime}x,5} \right)\left[
    B^{V}_{1}(m^2_{xy,\Xi},\mathcal{D}^\Xi_{YY},q)
    +B^{V}_{1}(m^2_{xy,\Xi},\mathcal{D}^\Xi_{XX},q) \right]\nonumber\\
  %%%%%%%%%%%%%%%%%%%%%%%%%%%%
  %%%%%%%%%%%%%%%%%%%%%%%%%%%%
  +&c_\Xi q_{\mu}\left[
    -B^{V}_{2\mu}(m^2_{xy,\Xi},\mathcal{D}^\Xi_{X^{\prime}Y},q)
    +B^{V}_{2\mu}(m^2_{xy,\Xi},\mathcal{D}^\Xi_{X^{\prime}X},q)\right]\nonumber\\
  %%%%%%%%%%%%%%%%%%%%%%%%%%%%
  %%%%%%%%%%%%%%%%%%%%%%%%%%%%
  +&\frac{1}{2}q_{\mu}\left[
    B^{V}_{2\mu}(m^2_{xy,\Xi},\mathcal{D}^\Xi_{YY},q)
    -B^{V}_{2\mu}(m^2_{xy,\Xi},\mathcal{D}^\Xi_{XX},q)\right]\nonumber\\
  %%%%%%%%%%%%%%%%%%%%%%%%%%%%
  %%%%%%%%%%%%%%%%%%%%%%%%%%%%
  +&\frac{1}{2}p_{12\mu}\left[
    B^{V}_{2\mu}(m^2_{xy,\Xi},\mathcal{D}^\Xi_{YY},q)
    -2B^{V}_{2\mu}(m^2_{xy,\Xi},\mathcal{D}^\Xi_{YX},q)\right.\nonumber\\
  &\hphantom{\frac{1}{2}p_{12\mu}}\left.
    +B^{V}_{2\mu}(m^2_{xy,\Xi},\mathcal{D}^\Xi_{XX},q)
    \right] \nonumber\\
  %%%%%%%%%%%%%%%%%%%%%%%%%%%%
  %%%%%%%%%%%%%%%%%%%%%%%%%%%%
  +& B^{V}_{}(m^2_{xy,\Xi},\mathcal{D}^\Xi_{X^{\prime}Y},q)\left(
    m^2_{x^{\prime}x^{\prime},5}
    -m^2_{x^{\prime}y,5}
    -m^2_{x^{\prime}x,5}c_\Xi
  \right) \nonumber\\
  %%%%%%%%%%%%%%%%%%%%%%%%%%%%
  %%%%%%%%%%%%%%%%%%%%%%%%%%%%
  +&B^{V}_{}(m^2_{xy,\Xi},\mathcal{D}^\Xi_{X^{\prime}X},q)\left(
    -m^2_{x^{\prime}x^{\prime},5}
    +m^2_{x^{\prime}y,5}c_\Xi
    +m^2_{x^{\prime}x,5}
  \right)
  \left.\vphantom{\frac{1}{2}}\right)\nonumber
\end{align}
\begin{align}
  \Delta^V h_\mu^{xy}= -\frac{1}{2f^2}\sum_\Xi\left(\vphantom{\frac{1}{2}}\right.
  %%%%%%%%%%%%%%%%%%%%%%%%%%%%
  %%%%%%%%%%%%%%%%%%%%%%%%%%%%
  &\frac{1}{4}\sum_{\mathcal{S}}\left[\vphantom{\frac[{1}{2}}
    -p_{12\nu}B^{V}_{23\mu\nu}(m^2_{x\mathcal{S},\Xi},m^2_{\mathcal{S}y,\Xi},q)\right.\\
  &\hphantom{\frac{1}{4}\sum_{\mathcal{S}}(}  +\frac{1}{2}B^{V}_{2\mu}(m^2_{x\mathcal{S},\Xi},m^2_{\mathcal{S}y,\Xi},q)\left(
    -q^2
    -m^2_{x^{\prime}y,5}
    +m^2_{x^{\prime}x,5}
  \right) \nonumber\\ 
  &\left.\hphantom{\frac{1}{4}\sum_{\mathcal{S}}(}
    -\frac{1}{2}\left\{2A^{V}_{\mu}(m^2_{x^{\prime}\mathcal{S},\Xi}) - A^{V}_{\mu}(m^2_{y\mathcal{S},\Xi}) -A^{V}_{\mu}(m^2_{x\mathcal{S},\Xi})\right\}\right]\nonumber\\
  % 
  %%%%%%%%%%%%%%%%%%%%%%%%%%%%
  %%%%%%%%%%%%%%%%%%%%%%%%%%%%
  %%%%%%%%%%%%%%%%%%%%%%%%%%%%
  -&p_{12\nu}\left[B^{V}_{23\mu\nu}(m^2_{xy,\Xi},\mathcal{D}^\Xi_{YY},q)
    -2B^{V}_{23\mu\nu}(m^2_{xy,\Xi},\mathcal{D}^\Xi_{YX},q)\right.\nonumber\\
  &\hphantom{p_{12\nu}}\left.+B^{V}_{23\mu\nu}(m^2_{xy,\Xi},\mathcal{D}^\Xi_{XX},q)
     \right] \nonumber\\
  %%%%%%%%%%%%%%%%%%%%%%%%%%%%
  %%%%%%%%%%%%%%%%%%%%%%%%%%%%
  +&B^{V}_{2\mu}(m^2_{xy,\Xi},\mathcal{D}^\Xi_{X^{\prime}Y},q)\left(
    -m^2_{x^{\prime}x^{\prime},5}
    +m^2_{x^{\prime}y,5}(1+c_\Xi)
    +m^2_{x^{\prime}x,5}c_\Xi
    +q^2c_\Xi
  \right) \nonumber\\
  %%%%%%%%%%%%%%%%%%%%%%%%%%%%
  %%%%%%%%%%%%%%%%%%%%%%%%%%%%
  +&B^{V}_{2\mu}(m^2_{xy,\Xi},\mathcal{D}^\Xi_{X^{\prime}X},q)\left(
    m^2_{x^{\prime}x^{\prime},5}
    -m^2_{x^{\prime}y,5}c_\Xi
    -m^2_{x^{\prime}x,5}(1+c_\Xi)
    -q^2c_\Xi
  \right) \nonumber\\
  %%%%%%%%%%%%%%%%%%%%%%%%%%%%
  %%%%%%%%%%%%%%%%%%%%%%%%%%%%
  +&\frac{1}{2}B^{V}_{2\mu}(m^2_{xy,\Xi},\mathcal{D}^\Xi_{YY},q)\left(
    -q^2
    -m^2_{x^{\prime}y,5}
    +m^2_{x^{\prime}x,5}
  \right) \nonumber\\
  %%%%%%%%%%%%%%%%%%%%%%%%%%%%
  %%%%%%%%%%%%%%%%%%%%%%%%%%%%
  +&\frac{1}{2}B^{V}_{2\mu}(m^2_{xy,\Xi},\mathcal{D}^\Xi_{XX},q)\left(
    +q^2
    -m^2_{x^{\prime}y,5}
    +m^2_{x^{\prime}x,5}
  \right) 
  \left.\vphantom{\frac{1}{2}}\right)\nonumber
\end{align}

\subsubsection{Finite-volume corrections for the scalar form factor}
\label{sec:rSPQscal}

\begin{align}
  \frac{\Delta^V\rho_{xy}}{m_K^2-m_\pi^2} = -\frac{1}{2f^2}\left(\vphantom{\frac{1}{2}}\right.
  %%%%%%%%%%%%%%%%%%%%%%%%%%%%
  %%%%%%%%%%%%%%%%%%%%%%%%%%%%
  &\frac{1}{8}\sum_{\mathcal{S}}\left[\vphantom{\frac{1}{2}}
-\left(m^2_{x^\prime y,5}-m^2_{x^\prime x,5}\right)B^{V}_{1}(m^2_{x\mathcal{S},\Xi},m^2_{\mathcal{S}y,\Xi},q)\right.\\
  &\hphantom{\frac{1}{8}\sum_{\mathcal{S}}(}+p_{12\mu}B^{V}_{2\mu}(m^2_{x\mathcal{S},\Xi},m^2_{\mathcal{S}y,\Xi},q)\nonumber\\
  &\left.\hphantom{\frac{1}{8}\sum_{\mathcal{S}}(}+\frac{1}{2}B^{V}_{}(m^2_{x\mathcal{S},\Xi},m^2_{\mathcal{S}y,\Xi},q)\left(
      q^2
      +m^2_{x^\prime y,5}
      -m^2_{x^\prime x,5}
    \right) \right]\nonumber\\
  %%%%%%%%%%%%%%%%%%%%%%%%%%%% 
  %%%%%%%%%%%%%%%%%%%%%%%%%%%% 
  +&\frac{1}{2}\left(m^2_{x^\prime y,5}-m^2_{x^\prime x,5}\right)\left[
    -B^{V}_{1}(m^2_{xy,\Xi},\mathcal{D}^\Xi_{YY},q)\right.\nonumber\\
  &\left.\hphantom{\frac{1}{2}\left(m^2_{x^\prime y,5}-m^2_{x^\prime x,5}\right)}
  +B^{V}_{1}(m^2_{xy,\Xi},\mathcal{D}^\Xi_{XX},q)\right]
 \nonumber\\
 %%%%%%%%%%%%%%%%%%%%%%%%%%%% 
  %%%%%%%%%%%%%%%%%%%%%%%%%%%%
  +&p_{12\mu}\frac{1}{2}\left[
    B^{V}_{2\mu}(m^2_{xy,\Xi},\mathcal{D}^\Xi_{YY},q)
    -B^{V}_{2\mu}(m^2_{xy,\Xi},\mathcal{D}^\Xi_{XX},q)  \right] \nonumber\\
  %%%%%%%%%%%%%%%%%%%%%%%%%%%%
  %%%%%%%%%%%%%%%%%%%%%%%%%%%%
  -& \frac{c_\Xi}{2} B^{V}_{}(m^2_{xy,\Xi},\mathcal{D}^\Xi_{X^{\prime}Y},q)\left(
    m^2_{x^{\prime} y,5}
    +m^2_{x^\prime x,5}
    +q^2
  \right) \nonumber\\
  %%%%%%%%%%%%%%%%%%%%%%%%%%%%
  %%%%%%%%%%%%%%%%%%%%%%%%%%%%
  -& \frac{c_\Xi}{2} B^{V}_{}(m^2_{xy,\Xi},\mathcal{D}^\Xi_{X^{\prime}X},q)\left(
    m^2_{x^{\prime} y,5}
    +m^2_{x^\prime x,5}
    +q^2
  \right) \nonumber\\
  %%%%%%%%%%%%%%%%%%%%%%%%%%%%
  %%%%%%%%%%%%%%%%%%%%%%%%%%%%
  +&\frac{1}{4}B^{V}_{}(m^2_{xy,\Xi},\mathcal{D}^\Xi_{YY},q)\left(
    m^2_{x^\prime y,5}
    -m^2_{x^\prime x,5}
    +q^2
  \right) \nonumber\\
  %%%%%%%%%%%%%%%%%%%%%%%%%%%%
  %%%%%%%%%%%%%%%%%%%%%%%%%%%%
  +&\frac{1}{2}B^{V}_{}(m^2_{xy,\Xi},\mathcal{D}^\Xi_{YX},q)q^2 \nonumber\\
  %%%%%%%%%%%%%%%%%%%%%%%%%%%%
  %%%%%%%%%%%%%%%%%%%%%%%%%%%%
  +&\frac{1}{4}B^{V}_{}(m^2_{xy,\Xi},\mathcal{D}^\Xi_{XX},q)\left(
    -m^2_{x^\prime y,5}
    +m^2_{x^\prime x,5}
    +q^2
  \right) \nonumber\\
  %%%%%%%%%%%%%%%%%%%%%%%%%%%%
  %%%%%%%%%%%%%%%%%%%%%%%%%%%%
  -&\frac{c_\Xi}{2}\left[A^{V}_{}(\mathcal{D}^\Xi_{X^{\prime}Y})+A^{V}_{}(\mathcal{D}^\Xi_{X^{\prime}X}) \right]
  \left.\vphantom{\frac{1}{2}}\right)\nonumber
\end{align}

\section{Typical finite-volume corrections to current lattice simulations}
\label{sec:numerical}

As an illustration of the numerical size of finite-volume corrections in
current lattice simulations, we present an explicit calculation of these effects
for the set of ensembles used by the FNAL/MILC collaboration in its on-going
analysis of $K\to\pi\ell\nu$ \cite{Gamiz:2016bpm}. The formulas in the previous section are of course more general.

The ensemble parameters we use are presented in Tables \ref{tab:ensembles}
and \ref{tab:splittings}. With the exception of the meson masses and the hairpin couplings, those parameters are originally listed in Ref.~\cite{Bazavov:2012xda}. 
The taste splittings shown in Table~\ref{tab:splittings} are averages over tastes that are degenerate under the residual $SO(4)$ taste symmetry, which is unbroken at the order to which we are working. 
We use the values of the relative scale $r_1/a$ together with the value of the absolute scale $r_1=0.3117(22)$~fm~\cite{Bazavov:2011aa} to convert lattice quantities, proportional to $a$, to physical units. The masses, originally determined in terms of $a$, are from the ongoing FNAL/MILC $K_{l3}$ analysis~\cite{Ktopilnu2016}. The hairpin couplings, $a^2\delta_V$ and $a^2\delta_A$ are from
an unpublished FNAL/MILC analysis for the $0.12$~fm lattice and have been scaled
by $\alpha_s^2a^2$ for the other cases.
Finally, we use $f=130.41$~MeV.

The numerical evaluations needed will be implemented
in CHIRON \cite{CHIRON}.

\begin{table}[t]
\caption{Parameters for the HISQ $N_f=2+1+1$ MILC ensembles we have used in
the numerical results~\cite{Bazavov:2012xda}.
The numbers not in that reference come from the on-going FNAL/MILC
analysis~\cite{Ktopilnu2016}. 
The light (up,down) valence quark masses are the same as the light sea quark
masses on each ensemble, but the strange quark can be different.
We have quoted the kaon mass therefore with valence and with sea quarks.
Below we refer to the different ensembles using $m_\pi$ and $m_\pi L$ since
these are the most relevant quantities in the finite-volume calculation.}
\label{tab:ensembles}
\begin{center}
\begin{tabular}{ccclcccc}
    \hline
    a & $m_l/m_s$ & L & $r_1/a$ & $m_\pi$ & $m_K$ & $m_K$(sea) & $m_\pi L$\\
 (fm) & & (fm) &  & (MeV) & (MeV) & (MeV) & \\
    \hline
    0.15 & 0.035 & 4.8 & 2.089  & 134 & 505 & 490 & 3.2\\
    \hline
    0.12 & 0.2   & 2.9 & 2.575  & 309 & 539 & 528 & 4.5\\
         & 0.1   & 2.9 & 2.5962 & 220 & 516 & 506 & 3.2\\
         & 0.1   & 3.8 & 2.5962 & 220 & 516 & 506 & 4.3\\
         & 0.1   & 4.8 & 2.5962 & 220 & 516 & 506 & 5.4\\
         & 0.035 & 5.7 & 2.608  & 135 & 504 & 493 & 3.9\\
    \hline
    0.09 & 0.2   & 2.9 & 3.499  & 312 & 539 & 534 & 4.5\\
         & 0.1   & 4.2 & 3.566  & 222 & 523 & 512 & 4.7\\
         & 0.035 & 5.6 & 3.565  & 129 & 495 & 495 & 3.7\\
    \hline
    0.06 & 0.2   & 2.8 & 5.342  & 319 & 547 & 547 & 4.5\\
         & 0.035 & 5.5 & 5.4424 & 134 & 491 & 491 & 3.7\\
    \hline
  \end{tabular}
\end{center}
\end{table}

\begin{table}[tbh]
\caption{Taste splittings and hairpin couplings for the HISQ $N_f=2+1+1$ MILC
ensembles we have used in the numerical
results. Taste splittings from Ref.~\cite{Bazavov:2012xda,hairpins}
and the $r_1/a$ in Table~\ref{tab:ensembles}
and hairpin vertices from an unpublished MILC analysis. 
The correspondence between ensembles here and in Table~\ref{tab:ensembles} is
given by the lattice spacing $a$ since the splittings used are the same for
all ensembles with a given lattice spacing.}
\label{tab:splittings}
\begin{center}
  \begin{tabular}{ccccccc}
    \hline
    $a$ & $a^2\delta_V$ & $a^2\delta_A$ & $a^2\Delta_V$ & $a^2\Delta_A$ &$a^2\Delta_T$ &$a^2\Delta_S$ \\
    (fm) & GeV$^2$ & GeV$^2$ & GeV$^2$ & GeV$^2$ & GeV$^2$ &  GeV$^2$ \\ 
    \hline
    0.15 & 0.042256 & -0.058008 & 0.11464 & 0.041394 & 0.077496 & 0.1474\\
    \hline
    0.12 & 0.022844 & -0.031341 & 0.062249 & 0.021744 & 0.041057 & 0.08288\\
    \hline
    0.09 & 0.0073091 & -0.010034 & 0.019641 & 0.0072139 & 0.01334 & 0.025289\\
    \hline
    0.06 & 0.0013934 & -0.0019131 & 0.003647 & 0.0013226 & 0.0024848 & 0.0051299\\
    \hline
  \end{tabular}
\end{center}
\end{table}

Next, we have to make a choice of which masses to use. From the
pion and kaon masses in Table~\ref{tab:ensembles}
we fix the lowest order masses\footnote{Corrections
are higher order than we have used in \chpt.} for the 
neutral particles (pseudo-scalar taste for staggered) via
\begin{align}
\label{massesinput}
m^2_{11}=\,&m^2_{22}=m^2_U=m^2_D=m_\pi^2
\nonumber\\
m^2_{33} =\,& 2 m_K^2-m_\pi^2 &
m^2_S =\,&  2 m_K^2(\mathrm{sea})-m_\pi^2\,,
\end{align}
where $m_{11}$, $m_{22}$ and $m_{33}$ are the masses of the neutral valence-valence mesons, and $m_U$, $m_D$ and $m_S$ are masses of the neutral sea-sea mesons. 
In the staggered theory we can determine the meson masses at LO in ChPT
using the relation
\begin{align}
  \label{sec:LOmass}
  m^2_{ab,\Xi}=\frac{1}{2}\left( m^2_{aa}+m^2_{bb}\right) + a^2\Delta_\Xi\,.
\end{align}
Alternatively we could have determined $m_{ss}^2$ and $m_S^2$
from the neutral meson masses obtained from the lattice instead of from the
kaon masses. We have checked that these two choices for the meson masses
produces differences which are small,
much below the expected size of higher orders of about 20\%.
All results presented here are calculated using the LO SChPT expression
in Eq.~(\ref{sec:LOmass}), together with the values for masses and
taste splittings in Tables 1 and 2 and Eq.~(\ref{massesinput}).

The finite-volume correction to $K_{l3}$ decays is presented in a way that
shows the relative size to the leading order, $f_+(0)^{LO}=1$.
We calculate each term in the Ward identity in Eq.~(\ref{eq:ward}) divided by
the mass difference,
\begin{align}
\label{eq:wardDivide}
\frac{\Delta^V m_K^2-\Delta^V m_\pi^2}{m_K^2-m_\pi^2}+\Delta^V f_+(0)
+\frac{-q_\mu h_\mu}{m_K^2-m_\pi^2} = \frac{\Delta^V\rho}{m_K^2-m_\pi^2},
\end{align}
at $q^2=0$. The results are presented in
Tables~\ref{tab:results1}-\ref{tab:results3}. The needed twist angle 
is determined by having $q^2=0$. While our analytical results are for a 
fully general twisting, the numerical examples presented here are for the case 
where we only twist the valence up quark. 
This corresponds to a kaon at rest and a moving pion.
We present results for three situations. 
Two are for rooted staggered quarks, with $\theta^u=(0,\theta,\theta,\theta)$,
Table~\ref{tab:results1}, and with $\theta^u=(0,\theta^\prime,0,0)$,
Table~\ref{tab:results2}. $\theta$ and $\theta^\prime$
are chosen to have $q^2=0$. The third set of results is for continuum 
quarks (no effects of staggering), with $\theta^u=(0,\theta^\prime,0,0)$,
Table~\ref{tab:results3}.

\begin{table}[t]
\caption{Values for the different parts in the Ward identity
in Eq.~(\ref{eq:wardDivide}) for the ensembles in Table \ref{tab:ensembles}.
The labels ``mass," ``$f_+$'' and ``$h_\mu$'' refer to the three terms 
in the left-hand side of Eq.~(\ref{eq:wardDivide}), and ``$\rho$,'' the
 right-hand side.
The numbers are obtained with $\theta^u=(0,\theta,\theta,\theta)$
such that $q^2=0$ and the kaon at rest. The effects due to staggered 
quarks are included.}
\label{tab:results1}
\begin{center}
\begin{tabular}{ccrrrr}
\hline
 $m_\pi$ & $m_\pi L$ & ``mass''~ & ``$f_+$''~~ & ``$h_\mu$''~~ & ``$\rho$''~~~\\ 
\hline
 134 & 3.2 & $ 0.00000$&  $-0.00042$& $0.00007$& $-0.00036$\\
\hline                                    
 309 & 4.5 & $ 0.00013$&  $-0.00003$&$-0.00041$& $-0.00031$ \\
 220 & 3.2 & $ 0.00054$&  $-0.00048$&$-0.00084$& $-0.00077$ \\
 220 & 4.3 & $-0.00007$&  $-0.00009$&$-0.00005$& $-0.00021$ \\
 220 & 5.4 & $-0.00005$&  $-0.00003$& $0.00001$& $-0.00006$ \\
 135 & 3.9 & $-0.00006$&  $-0.00020$& $0.00005$& $-0.00021$ \\
\hline                                                  
 312 & 4.5 & $ 0.00047$&   $0.00023$&$-0.00068$& $-0.00001$ \\
 222 & 4.7 & $-0.00000$&  $ 0.00018$&$-0.00003$& $ 0.00014$ \\
 129 & 3.7 & $-0.00013$&  $-0.00004$&$ 0.00009$& $-0.00007$ \\
\hline                                                  
 319 & 4.5 & $ 0.00052$&   $0.00037$&$-0.00081$& $ 0.00008$ \\
 134 & 3.7 & $-0.00016$&   $0.00045$& $0.00013$& $ 0.00043$ \\
\hline
\end{tabular}
\end{center}
\end{table}

\begin{table}[tbh]
\caption{Values for the different parts in the Ward identity
in Eq.~(\ref{eq:wardDivide}) for the ensembles in Table \ref{tab:ensembles}.
The labels ``mass," ``$f_+$'' and ``$h_\mu$'' refer to the three terms 
in the left-hand side of Eq.~(\ref{eq:wardDivide}), and ``$\rho$,'' the
 right-hand side.
The numbers are obtained with $\theta^u=(0,\theta',0,0)$
such that $q^2=0$ and the kaon at rest. The effects due to staggered 
quarks are included.}
\label{tab:results2}
\begin{center}
\begin{tabular}{ccrrrr}
\hline
 $m_\pi$ & $m_\pi L$ & ``mass''~ & ``$f_+$''~~ & ``$h_\mu$''~~ & ``$\rho$''~~~\\ 
\hline
 134 & 3.2 & $-0.00003$&  $-0.00066$& $0.00008$& $-0.00061$\\
\hline                                    
 309 & 4.5 & $-0.00030$&  $-0.00017$&$-0.00002$& $-0.00049$ \\
 220 & 3.2 & $-0.00078$&  $-0.00105$& $0.00036$& $-0.00148$ \\
 220 & 4.3 & $-0.00033$&  $-0.00034$& $0.00018$& $-0.00049$ \\
 220 & 5.4 & $-0.00008$&  $-0.00010$& $0.00003$& $-0.00015$ \\
 135 & 3.9 & $-0.00002$&  $-0.00032$& $0.00001$& $-0.00033$ \\
\hline                                                  
 312 & 4.5 & $-0.00019$&   $0.00002$&$-0.00009$& $-0.00026$ \\
 222 & 4.7 & $-0.00024$&  $-0.00018$& $0.00017$& $-0.00025$ \\
 129 & 3.7 & $-0.00003$&  $-0.00050$&$-0.00001$& $-0.00054$ \\
\hline                                                  
 319 & 4.5 & $-0.00026$&   $0.00013$&$-0.00012$& $-0.00025$ \\
 134 & 3.7 & $-0.00005$&  $-0.00058$& $0.00001$& $-0.00062$ \\
\hline
\end{tabular}
\end{center}
\end{table}

\begin{table}[tbh]
\caption{Values for the different parts in the Ward identity
in Eq.~(\ref{eq:wardDivide}) for the ensembles in Table \ref{tab:ensembles}.
The labels ``mass," ``$f_+$'' and ``$h_\mu$'' refer to the three terms 
in the left-hand side of Eq.~(\ref{eq:wardDivide}), and ``$\rho$,'' the
 right-hand side.
The numbers are obtained with $\theta^u=(0,\theta',0,0)$
such that $q^2=0$ and the kaon at rest. This is the case without effects
from staggering.}
\label{tab:results3}
\begin{center}
\begin{tabular}{ccrrrr}
\hline
 $m_\pi$ & $m_\pi L$ & ``mass''~ & ``$f_+$''~~ & ``$h_\mu$''~~ & ``$\rho$''~~~\\ 
\hline
 134 & 3.2 & $-0.00049$&  $-0.00124$& $0.00037$& $-0.00137$\\
\hline                                    
 309 & 4.5 & $-0.00033$&  $ 0.00014$&$-0.00004$& $ 0.00022$ \\
 220 & 3.2 & $-0.00113$&  $ 0.00077$&$ 0.00067$& $ 0.00031$ \\
 220 & 4.3 & $-0.00062$&  $-0.00011$& $0.00046$& $-0.00027$ \\
 220 & 5.4 & $-0.00014$&  $-0.00011$& $0.00010$& $-0.00016$ \\
 135 & 3.9 & $ 0.00004$&  $-0.00045$&$-0.00008$& $-0.00049$ \\
\hline                                                  
 312 & 4.5 & $ 0.00031$&  $ 0.00015$&$-0.00009$& $-0.00025$ \\
 222 & 4.7 & $-0.00037$&  $-0.00015$& $0.00027$& $-0.00025$ \\
 129 & 3.7 & $-0.00000$&  $-0.00066$&$-0.00005$& $-0.00071$ \\
\hline                                                  
 319 & 4.5 & $-0.00031$&   $0.00015$&$-0.00011$& $-0.00027$ \\
 134 & 3.7 & $-0.00007$&  $-0.00064$& $0.00001$& $-0.00070$ \\
\hline
\end{tabular}
\end{center}
\end{table}

Looking at the tables one effect is very clear: For these lattices the 
finite-volume corrections are all very small and clearly below the 0.2\% used
as error in the published FNAL/MILC results~\cite{Bazavov:2013maa}.
The finite-volume effects also decrease with increasing $m_\pi L$
as expected.

A second observation is that the finite-volume effects are dependent on the
precise way the twisting is done. The predictions for twisting in
all space directions or in one space direction only are quite different, 
as a glance at Tables~\ref{tab:results1} and \ref{tab:results2} shows.
This suggests a relatively cheap way to check the rough size of 
finite-volume effects, as well as our predictions for them:  Perform the 
(lattice) calculations with different choices for the partial twisting but 
using the same underlying lattice.

A third observation is that the finite-volume correction is typically
smaller for the case with staggered effects than for the
continuum case. The differences can be of the same size as the actual
corrections. We believe this is due to the fact that the non-pseudoscalar
taste mesons have typically larger masses and thus have smaller 
finite-volume effects.

\begin{figure}[h!t]
  \label{fig:ward}
  \centering
  \includegraphics{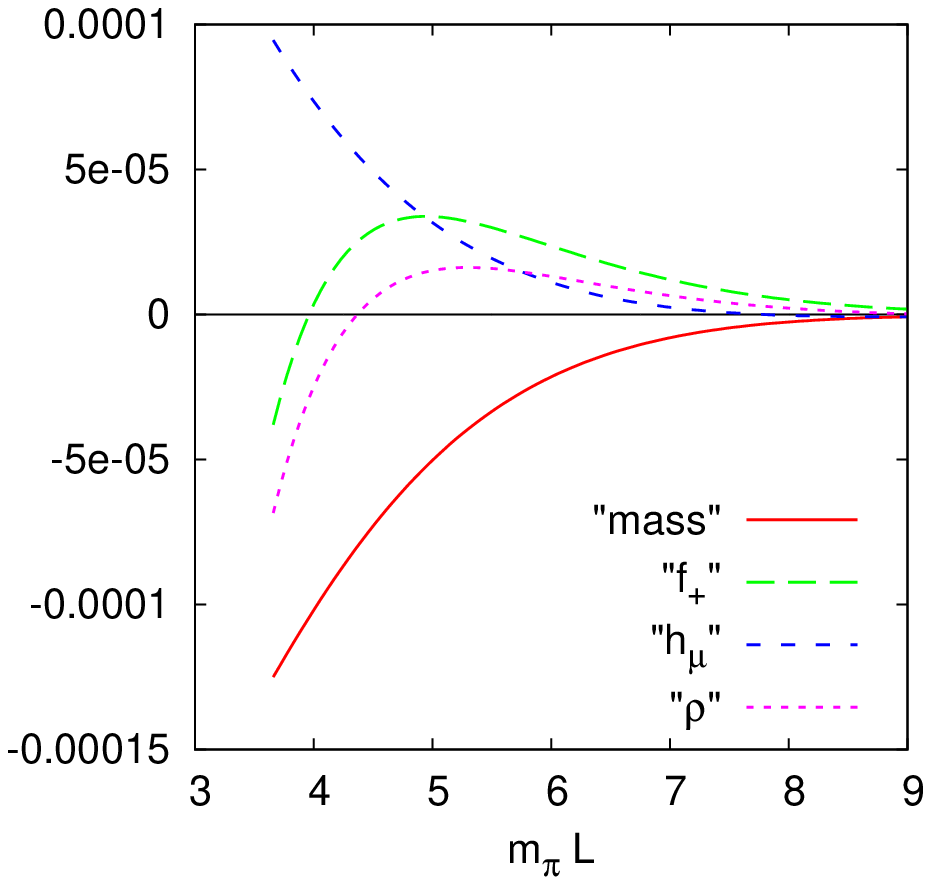}
  \caption{Values for the different parts in the Ward identity in
 Eq.~(\ref{eq:wardDivide}) when varying the pion mass 
 while keeping the kaon mass fixed with the staggered parameters from the
 ensemble with $m_\pi=129$ and $m_\pi L= 3.7$ in Table \ref{tab:ensembles}.
The labels ``mass'', ``$f_+$'' and ``$h_\mu$`` refer to the three terms 
in the left-hand side of Eq.~(\ref{eq:wardDivide}), and ``$\rho$'',  
the right-hand side. The numbers are obtained with 
$\theta^u=(0,\theta,\theta,\theta)$ such that $q^2=0$ and the kaon at rest.}
\end{figure}

The exponential decrease of the finite-volume correction
with $m_\pi L$ remains valid here. As an example, Fig.~\ref{fig:ward} shows 
the contributions to Eq.~(\ref{eq:wardDivide}) as a function of $m_\pi L$.  
We have used the parameters of the ensemble with $m_\pi= 129$~MeV and 
$m_\pi L=3.7$. We then vary $m_\pi$ while keeping the valence and sea kaon 
masses fixed.

\section{Conclusions}
\label{sec:conclusions}

In this paper we have calculated the finite-volume corrections to $K_{l3}$
decays in rooted staggered partially-quenched {\chpt} with twisted boundary
conditions. We allow for different twists in the valence and sea sector as well.
The analytical formulas in section~\ref{sec:analytical} and the appendices
are our main results. By replacing the finite-volume correction functions 
with their infinite-volume counterparts, these formulas can also be used 
to obtain the corresponding infinite-volume expressions. We have 
presented results for the vector as well as the scalar form factor. 
We have checked analytically and numerically that the relevant Ward identity
is fulfilled. 

Numerically, for representative parameters of current lattice simulations,
the corrections may be as large as $\mathcal{O}(10^{-3})$, but are often much 
smaller. The magnitude and sign of the corrections vary significantly 
between ensembles.

As a relatively cheap way to check for finite-volume effects, we suggest 
comparing the results for different partial twist choices on the same 
underlying configurations.

\section*{Acknowledgments}

This work is supported in part by the Swedish Research Council grants
contract numbers 621-2013-4287 and 2015-04089, by
the European Research Council (ERC) under the European Union's Horizon 2020
research and innovation program (grant agreement No 668679), 
by MINECO under grant number FPA2013-47836-C-1-P,
by Junta de Andaluc\'{\i}a grants FQM 101 and FQM 6552,
and by the U.S. Department of Energy under Grant DE-FG02-91ER-40628.
\appendix
\section{Integrals and relations}
\label{sec:integrals}

Our results can be written using slight modifications of integrals found elsewhere in the literature. In this section we define the integrals we need and give references to where more detailed treatments can be found.

\subsection{One loop integrals with single poles}

We will use the notation for finite-volume integrals given in Eq.~(\ref{eq:twistInt}). Note that every integral below depends on the twist angles since these determine which momenta are sampled in the sum in Eq.~(\ref{eq:twistInt}). We use the mass in a given propagator to indicate which momenta are to be sampled in each integral. For example a momentum $k$ which shows up as $(k^2+m_{\pi^+}^2)$ will only assume the allowed values for a $\pi^+$ meson. For this reason, $(q-k)^2+m_2^2$ is not equivalent $(k-q)^2+m_2^2$  because the former case implies that $q-k$ takes the allowed values for the meson 2, while in the later case $q-k$ takes the allowed values for the antiparticle of meson 2. 

All our results are given in Euclidean space.  We need the following integrals
\begin{align}
  \mathcal{A}(m^2) &= -\int_V \frac{d^dk}{(2\pi)^d}\frac{1}{(k^2+m^2)}\\
  \mathcal{A}_\mu(m^2) &= -\int_V \frac{d^dk}{(2\pi)^d}\frac{k_\mu}{(k^2+m^2)}\nonumber\\
  \mathcal{B}(m_1^2,m_2^2,q) &= \int_V \frac{d^dk}{(2\pi)^d}\frac{1}{(k^2+m_1^2)((q-k)^2+m_2^2,q)}\nonumber\\
  \mathcal{B}_\mu(m_1^2,m_2^2,q) &= \int_V \frac{d^dk}{(2\pi)^d}\frac{k_\mu}{(k^2+m_1^2)((q-k)^2+m_2^2,q)}\nonumber\\
  \mathcal{B}_{\mu\nu}(m_1^2,m_2^2,q) &= \int_V \frac{d^dk}{(2\pi)^d}\frac{k_\mu k_\nu}{(k^2+m_1^2)((q-k)^2+m_2^2,q)}\nonumber
\end{align}
We split these integrals according to
\begin{align}
  \mathcal{B}_\mu(m_1^2,m_2^2,q) &= q_\mu \mathcal{C}_1(m_1^2,m_2^2,q) + \mathcal{C}_{2\mu}(m_1^2,m_2^2,q,q)\\
  \mathcal{B}_{\mu\nu}(m_1^2,m_2^2,q) &= q_\mu q_\nu \mathcal{C}_{21}(m_1^2,m_2^2,q,q) \nonumber\\
  &- \delta_{\mu\nu}\mathcal{C}_{22}(m_1^2,m_2^2,q) + \mathcal{C}_{23\mu\nu}(m_1^2,m_2^2,q).\nonumber
\end{align}
where $\mathcal{C}_{2\mu}$ and $\mathcal{C}_{23\mu\nu}$ are zero due to symmetry in the zero twist and infinite-volume cases. The sign of $\mathcal{C}_{22}$ is chosen such that the corresponding Minkowski integral has plus signs for all three terms.

In this paper we are primarily interested in the finite volume part of the integrals. We denote the difference between the integral in finite and infinite volume by 

\begin{align}
  \mathcal{C}_{x} &\rightarrow B^V_{x},\\
  \mathcal{A}_{x} &\rightarrow A^V_{x}\nonumber
\end{align}
Expressions for these integrals in terms of Jacobi theta functions can be found in 
Ref.~\cite{Bijnens:2014yya}.

\subsection{One loop integrals for diagonal propagator}

In partially-quenched and staggered {\chpt} the diagonal propagators are more complicated than in standard {\chpt}, see sections \ref{sec:PQ} and \ref{sec:staggered}. The quark-flow disconnected part of the propagators takes the generic form
\begin{align}
  \mathcal{D}_{XY} = -\delta\frac{\prod_{i\in U,D,S}(p^2+m_i^2)}{(p^2+m_X^2)(p^2+m_Y^2)\prod_{j\in\pi^0,\eta,\eta^\prime}(p^2+m_j^2)}
\end{align}
where $\delta$ is the hairpin coupling of the propagating particles. In the staggered theory $\delta$ is taste dependent and given by
\begin{equation}
  \delta_\Xi = \
  \begin{cases} 
    a^2 \delta_V \equiv 16 a^2(C_{2V}-C_{5V})/f^2, &\Xi\in\{\xi_\mu\}\ {\rm (vector\ taste);}\cr
    a^2 \delta_A \equiv 16 a^2(C_{2A}-C_{5A})/f^2, &\Xi\in\{\xi_5\xi_\mu\}\ {\rm (axial\ taste);}\cr
    4m_0^2/3, &\Xi=I\ {\rm (singlet\ taste);}\cr 
    0, &{\rm otherwise.}\cr 
  \end{cases}
\end{equation}
The coefficients $C_{2V},\ldots$ are part of the taste breaking
potential $\mathcal{V}$ and are defined in Ref.~\cite{Aubin:2003mg}.
In the partially-quenched theory without taste, $\delta$ is given by
\begin{align}
  \delta = m_0^2/3.
\end{align}

Taking the isospin limit for the sea quarks, the diagonal propagators simplify to
\begin{align}
  \mathcal{D}_{XY} = -\delta\frac{\prod_{i\in U,S}(p^2+m_i^2)}{(p^2+m_X^2)(p^2+m_Y^2)\prod_{j\in\eta,\eta^\prime}(p^2+m_j^2)}.
\end{align}

To evaluate integrals with diagonal propagators we use the residue notation described in Ref.~\cite{Aubin:2003mg}. Both single and double poles can be evaluated using this technique. Double poles are written as derivatives of single poles. Although this method works well for evaluation, it produces rather messy expressions. For this reason we use generalized notation in which any of the $m^2$ arguments in the integrals may be replaced by $\mathcal{D}$, as in Ref.~\cite{Bernard:2013eya}. An example would be
\begin{align}
  &A^V(\mathcal{D}_{XY}^A) = -\int_V\frac{d^dk}{(2\pi)^d}\left(-a^2 \delta_{A}\right)\times\\
  &\left(\frac
  {(p^2+m_{U,A})(p^2+m_{D,A})(p^2+m_{S,A})}
  {(p^2+m_{X,A})(p^2+m_{Y,A})(p^2+m_{\pi^0,A})(p^2+m_{\eta,A})(p^2+m_{\eta^\prime,A})}\right)\nonumber.
\end{align}

\subsection{Integral relations}
There are relations among the integrals presented above.
The relations valid when including twisted boundary conditions
can be found in Ref.~\cite{Bijnens:2014yya}. In addition there are some relations which are useful for the neutral propagator given in Ref.~\cite{Aubin:2003mg}. Finally, we have used the relation
\begin{align}
  (m_a^2-m_b^2)\mathcal{D}_{ab} + (m_b^2-m_c^2)\mathcal{D}_{bc} + (m_c^2-m_a^2)\mathcal{D}_{ac} = 0.
\end{align}
All of these relations are needed to get the results in the forms presented above,  and they are necessary to show that the Ward identity is fulfilled.

There is also another class of relations among the integrals. These come from interchanging the masses in $\tilde B$ type integrals, which corresponds to changing the routings in the corresponding Feynman diagrams. The interchanges give the following behavior
\begin{align}
  B(m_1^2,m_2^2,q) &= B(m_2^2,m_1^2,q)\\
  B_1(m_1^2,m_2^2,q) &= B(m_2^2,m_1^2,q) - B_1(m_2^2,m_1^2,q)\nonumber\\
  B_{2\mu}(m_1^2,m_2^2,q) &= -B_{2\mu}(m_2^2,m_1^2,q)\nonumber\\
  B_{21}(m_1^2,m_2^2,q) &= B(m_2^2,m_1^2,q) - 2B_1(m_2^2,m_1^2,q) + B_{21}(m_2^2,m_1^2,q)\nonumber\\
  B_{22}(m_1^2,m_2^2,q) &= B_{22}(m_2^2,m_1^2,q)\nonumber\\
  B_{23\mu\nu}(m_1^2,m_2^2,q) &= B_{23\mu\nu}(m_2^2,m_1^2,q)\nonumber\\ 
  &- q_\mu B_{2\nu}(m_2^2,m_1^2,q) - q_\nu B_{2\nu}(m_2^2,m_1^2,q).\nonumber 
\end{align}
The last of these relations shows that the split between $f_-$ and $h_\mu$ is not unique.

All of the relations presented in this section are valid in both finite and infinite volume. In infinite volume some of the integrals are zero.

\section{Finite-volume corrections for masses}
\label{sec:finiteVolumeMasses}

In this appendix we give expressions for the finite-volume correction for masses in partially-quenched partially-twisted \chpt{} and partially-quenched partially-twisted rooted staggered \chpt. The expressions are valid for a flavor-charged meson with flavor content $xy$ and, in the staggered case, pseudoscalar taste. Note that in comparing with Ref.~\cite{Bijnens:2014yya} we see that the PQ expression neatly summarizes all results for flavor-charged mesons into a single expression, valid both with and without isospin.
\subsection{Partially-Quenched Partially-Twisted \chpt}

\begin{align}
  \Delta^Vm^2_{xy} = 
  -\frac{2}{f^2}\left(\sum_{\mathcal{S}}p_\mu\left[ A^{V}_{\mu}(m^2_{y\mathcal{S}}) - A^{V}_{\mu}(m^2_{x\mathcal{S}})\right]
    -m^2_{xy} A^V(\mathcal{D}_{XY})\right)
\end{align}

\subsection{Partially-Quenched Partially-Twisted Rooted Staggered \chpt}

\begin{align}
  \Delta^Vm^2_{xy,5} = 
  -\frac{1}{2f^2}\sum_\Xi\left(\sum_{\mathcal{S}}
      \frac{p_\mu}{4}\left[ A^{V}_{\mu}(m^2_{y\mathcal{S},\Xi}) - A^{V}_{\mu}(m^2_{x\mathcal{S},\Xi})\right]
    -m^2_{xy,5}A^V(\mathcal{D}_{XY}^\Xi)c_\Xi\right)
\end{align}

\section{$K^0 \rightarrow \pi^-$ isospin limit expressions}
\label{sec:isospinKl3}
In this appendix we present expressions for the process $K^0 \rightarrow \pi^-$ with up and down masses set equal; note that isospin is still broken by the boundary conditions. We give expressions for when sea and valence quarks have the same twist, which we call fully twisted, and for the partially-twisted case. In the partially-twisted case the indices $1,2,3$ on the masses indicate valence quarks $u,d,s$ respectively.
\subsection{Fully twisted}
\begin{align}
  \Delta^V f_+ = 
  -\frac{1}{2f^2}\left(\vphantom{\frac{1}{2}}\right.
    &4B_{22}^V(m_{\pi^+}^2,m_{K^0}^2,q) 
    + 6B_{22}^V(m_{K^+}^2,m_{\eta}^2,q)\\
    + &2B_{22}^V(m_{\pi^0}^2,m_{K^+}^2,q)
    - A^V(m_{\pi^+}^2)
    - 2A^V(m_{K^+}^2)\nonumber\\
    - &\left. A^V(m_{K^0}^2)
    - \frac{3}{2}A^V(m_{\eta}^2)
    - \frac{1}{2}A^V(m_{\pi^0}^2)\right)\nonumber
\end{align}
\begin{align}
  \Delta^V f_- =
  -\frac{1}{2f^2}\left(\vphantom{\frac{1}{2}}\right.
  &( m_{K}^2 - m_{\pi}^2)\left[4B_{21}^V(m_{\pi^+}^2,m_{K^0}^2,q) + 6B_{21}^V(m_{K^+}^2,m_{\eta}^2,q)\right.\\ 
   &\left.\hphantom{( m_{K}^2 - m_{\pi}^2)}+2B_{21}^V(m_{\pi^0}^2,m_{K^+}^2,q) - 4B_{1}^V(m_{\pi^+}^2,m_{K^0}^2,q)\right]\nonumber\\
  - &4B_{1}^V(m_{K^+}^2,m_{\eta}^2,q)(2m_{K}^2-m_{\pi}^2)\nonumber\\
  - &4m_{K}^2B_{1}^V(m_{\pi^0}^2,m_{K^+}^2,q)\nonumber\\
  + &2p_{K\mu}B_{2\mu}^{V}(m_{\pi^+}^2,m_{K^0}^2,q)\nonumber\\
  + &3p_{K\mu}B_{2\mu}^{V}(m_{K^+}^2,m_{\eta}^2,q)
  + 3p_{K\mu}B_{2\mu}^{V}(m_{\pi^0}^2,m_{K^+}^2,q)\nonumber\\
  + &2m_{K}^2B^V(m_{K^+}^2,m_{\eta}^2,q)
  + 2m_{\pi}^2B^V(m_{\pi^0}^2,m_{K^+}^2,q)\left.\vphantom{\frac{1}{2}},q\right)\nonumber
\end{align}
\begin{align}
\Delta^V h_\mu  = 
-\frac{1}{2f^2}\left(\vphantom{\frac{1}{2}}\right.
 -&4p_{12\nu}B_{23\mu\nu}^{V}(m_{\pi^+}^2,m_{K^0}^2,q)
 -6p_{12\nu}B_{23\mu\nu}^{V}(m_{K^+}^2,m_{\eta}^2,q)\\
 -&4p_{12\nu}B_{23\mu\nu}^{V}(m_{\pi^0}^2,m_{K^+}^2,q) + 2p_{12\nu}B_{23\mu\nu}^{V}(m_{K^+}^2,m_{\pi^0}^2,q)\nonumber\\
 + &2B_{2\mu}^{V}(m_{\pi^+}^2,m_{K^0}^2,q) (
 -q^2
 +m_{\pi}^2
 -m_{K}^2
 )\nonumber\\
 + &B_{2\mu}^{V}(m_{K^+}^2,m_{\eta}^2,q) (
 - 3q^2
 + m_{\pi}^2
 - 5m_{K}^2
 )\nonumber\\
 + &B_{2\mu}^{V}(m_{\pi^0}^2,m_{K^+}^2,q) (
 - 3q^2
 + m_{\pi}^2
 - 5m_{K}^2
 )\nonumber\\
 +6&\left(A^{V}_{\mu}(m_{\pi^+}^2) - A^{V}_{\mu}(m_{K^0}^2)\right)\left.\vphantom{\frac{1}{2}}\right)\nonumber
\end{align}

\begin{align}
  \frac{\Delta^V\rho_{xy}}{m_K^2-m_\pi^2} = 
  -\frac{1}{2f^2}  \left(\vphantom{\frac{1}{2}}\right.
  &-\left(m_{K}^2-m_{\pi}^2\right)\left[2B_{1}^{V}(m_{\pi^+}^2,m_{K^0}^2,q) \right.\\
  &\hphantom{-\left(m_{K}^2-m_{\pi}^2\right)}\left.+ B_{1}^{V}(m_{K^+}^2,m_{\eta}^2,q)+B_{1}^{V}(m_{\pi^0}^2,m_{K^+}^2,q)\right]\nonumber\\
  + &2p_{12\mu}B_{2\mu}^{V}(m_{\pi^+}^2,m_{K^0}^2,q)\nonumber\\
  + &p_{12\mu}B_{2\mu}^{V}(m_{K^+}^2,m_{\eta}^2,q)
  + p_{12\mu}B_{2\mu}^{V}(m_{\pi^0}^2,m_{K^+}^2,q)\nonumber\\
  + &B^V(m_{\pi^+}^2,m_{K^0}^2,q) (
  q^2
  -m_{\pi}^2
  +m_{K}^2
  )\nonumber\\
  + &\frac{1}{2}B^V(m_{K^+}^2,m_{\eta}^2,q) (
  + q^2
  - \frac{1}{3}m_{\pi}^2
  + \frac{5}{3}m_{K}^2
  )\nonumber\\
  + &\frac{1}{2}B^V(m_{\pi^0}^2,m_{K^+}^2,q) (
  3q^2
  + m_{\pi}^2
  + 3m_{K}^2
  )\nonumber\\
  + &\frac{1}{3}A^{V}(m_{\eta}^2)
  +A^{V}(m_{\pi^0}^2)\left.\vphantom{\frac{1}{2}}\right)\nonumber
\end{align}

\subsection{Partially twisted}

In the partially-twisted result there is no difference between sea and valence indices for flavor-neutral mesons. We label these states with $m_\pi$, $m_\eta$ and $m_S$ where $m_S=m_{33}$.

\begin{align}
  \Delta^V f_+ = 
  -\frac{1}{2f^2}\left(\vphantom{\frac{1}{2}}\right.
  &\sum_{\mathcal{S}} \left[- A^{V}(m_{1\mathcal{S}}^2) - A^{V}(m_{3\mathcal{S}}^2)
    + 4B_{22}^V(m_{1\mathcal{S}}^2,m_{\mathcal{S}3}^2,q)\right]\\
  &+6B_{22}^V(m_{13}^2,m_{\eta}^2,q)
  - 4B_{22}^V(m_{13}^2,m_{S}^2,q)\nonumber\\
  - &2B_{22}^V(m_{\pi^0}^2,m_{13}^2,q)\nonumber\\
  + &\frac{1}{2}\left[-3A^{V}(m_{\eta}^2)
    + A^{V}(m_{\pi^0}^2)
    + 2A^{V}(m_{S}^2)\right]\left.\vphantom{\frac{1}{2}}\right)\nonumber
\end{align}
\begin{align}
  \Delta^V f_- = 
  -\frac{1}{2f^2}\left(\vphantom{\frac{1}{2}}\right.
  &\sum_{\mathcal{S}} \left[
    -4(m_{K}^2 - m_{\pi}^2)\left\{
      -B_{21}^V(m_{1\mathcal{S}}^2,m_{\mathcal{S}3}^2,q) + B_{1}^V(m_{1\mathcal{S}}^2,m_{\mathcal{S}3}^2,q)\right\}\right.\nonumber\\
  &\hphantom{\sum_{\mathcal{S}} (}\left.+ 4p_{K\mu}B_{2\mu}^{V}(m_{1\mathcal{S}}^2,m_{\mathcal{S}3}^2,q)\right]\\
  + &(m_{K}^2-m_{\pi}^2)\left[6B_{21}^V(m_{13}^2,m_{\eta}^2,q)- 4B_{21}^V(m_{13}^2,m_{S}^2,q)\right.\nonumber\\
  &\left.\hphantom{(m_{K}^2-m_{\pi}^2)} - 2B_{21}^V(m_{\pi^0}^2,m_{13}^2,q)\right]\nonumber\\
  - &4B_{1}^V(m_{13}^2,m_{\eta}^2,q) (2m_{K}^2 - m_{\pi}^2)\nonumber\\
  + &4B_{1}^V(m_{13}^2,m_{S}^2,q) (m_{K}^2 - m_{\pi}^2)\nonumber\\
  - &4B_{1}^V(m_{\pi^0}^2,m_{13}^2,q)m_{\pi}^2\nonumber\\
  + &6p_{K\mu}B_{2\mu}^{V}(m_{13}^2,m_{\eta}^2,q)\nonumber\\
  - &4p_{K\mu}B_{2\mu}^{V}(m_{13}^2,m_{S}^2,q)
  + 2p_{K\mu}B_{2\mu}^{V}(m_{\pi^0}^2,m_{13}^2,q)\nonumber\\
  + &2B^V(m_{13}^2,m_{\eta}^2,q) m_{K}^2 + 2B^V(m_{\pi^0}^2,m_{13}^2,q) m_{\pi}^2
\left.\vphantom{\frac{1}{2}}\right)\nonumber
\end{align}
\begin{align}
  \Delta^V h_\mu = 
  -\frac{1}{2f^2}\left(\vphantom{\frac{1}{2}}\right.
  &\sum_{\mathcal{S}}\left[- 4p_{12\nu}B_{23\mu\nu}^{V}(m_{1\mathcal{S}}^2,m_{\mathcal{S}3}^2,q)\right.\\
  &\hphantom{\sum_{\mathcal{S}}(}+ 2B_{2\mu}^{V}(m_{1\mathcal{S}}^2,m_{\mathcal{S}3}^2,q) (
  - q^2 
  + m_{\pi}^2
  - m_{K}^2
  )\nonumber\\
  &\hphantom{\sum_{\mathcal{S}}(}+ 2 \left.\left\{A^{V}_{\mu}(m_{1\mathcal{S}}^2) -2A^{V}_{\mu}(m_{2\mathcal{S}}^2) + A^{V}_{\mu}(m_{3\mathcal{S}}^2)\right\}\right]\nonumber\\
  - &6p_{12\nu}B_{23\mu\nu}^{V}(m_{13}^2,m_{\eta}^2,q)
  + 4p_{12\nu}B_{23\mu\nu}^{V}(m_{13}^2,m_{S}^2,q)\nonumber\\
  + &2p_{12\nu}B_{23\mu\nu}^{V}(m_{13}^2,m_{\pi^0}^2,q)\nonumber\\
  + &B_{2\mu}^{V}(m_{13}^2,m_{\eta}^2,q) (
  - 3q^2 
  + m_{\pi}^2
  - 5m_{K}^2
  )\nonumber\\
  + &2B_{2\mu}^{V}(m_{13}^2,m_{S}^2,q)  (
  q^2 
  - m_{\pi}^2
  + m_{K}^2
  )\nonumber\\
  + &B_{2\mu}^{V}(m_{\pi^0}^2,m_{13}^2,q)  (
  - q^2 
  - m_{\pi}^2
  - 3m_{K}^2
  )
  \left.\vphantom{\frac{1}{2}}\right)\nonumber
\end{align}

\begin{align}
  \frac{\Delta^V\rho_{xy}}{m_K^2-m_\pi^2} = 
  -\frac{1}{2f^2}  \left(\vphantom{\frac{1}{2}}\right.
  &\sum_{\mathcal{S}}\left[- \left(m_{K}^2-m_{\pi}^2\right)2B_{1}^{V}(m_{1\mathcal{S}}^2,m_{\mathcal{S}3}^2,q)\right.\\
  &\hphantom{\sum_{\mathcal{S}}(}+ 2p_{12\mu}B_{2\mu}^{V}(m_{1\mathcal{S}}^2,m_{\mathcal{S}3}^2,q)\nonumber\\
  &\left.\hphantom{\sum_{\mathcal{S}}(}+ B^{V}(m_{1\mathcal{S}}^2,m_{\mathcal{S}3}^2,q)  (
  q^2
  - m_{\pi}^2
  + m_{K}^2
  )\right]\nonumber\\
+&\left(m_{K}^2-m_{\pi}^2\right)\left[-B_{1}^{V}(m_{13}^2,m_{\eta}^2,q)
  +2B_{1}^{V}(m_{13}^2,m_{S}^2,q)\right.\nonumber\\
  &\left.\hphantom{\left(m_{K}^2-m_{\pi}^2\right)}+B_{1}^{V}(m_{\pi^0}^2,m_{13}^2,q)\right]\nonumber\\
  +&p_{12\mu}B_{2\mu}^{V}(m_{13}^2,m_{\eta}^2,q)
  -2 p_{12\mu}B_{2\mu}^{V}(m_{13}^2,m_{S}^2,q)\nonumber\\
  - &p_{12\mu}B_{2\mu}^{V}(m_{\pi^0}^2,m_{13}^2,q)\nonumber\\
  + &\frac{1}{2}B^{V}(m_{13}^2,m_{\eta}^2,q) (
  q^2
  - \frac{1}{3}m_{\pi}^2
  + \frac{5}{3}m_{K}^2
  )\nonumber\\
  + &B^{V}(m_{13}^2,m_{S}^2,q)  (
  -q^2
  + m_{\pi}^2
  - m_{K}^2
  )\nonumber\\
  + &\frac{1}{2}B^{V}(m_{\pi^0}^2,m_{13}^2,q) (
  q^2
  + 3m_{\pi}^2
  + m_{K}^2
  )\nonumber\\
  + &\frac{1}{3}A^{V}(m_{\eta}^2)
  + A^{V}(m_{\pi^0}^2)\left.\vphantom{\frac{1}{2}}\right)\nonumber
\end{align}


\begin{thebibliography}{99}

\bibitem{Bazavov:2013maa}
 A.~Bazavov {\it et al.},
  %\emph{Determination of $|V_{us}|$ from a lattice-QCD calculation of the $K\to\pi\ell\nu$ semileptonic form factor with physical quark masses},
  Phys.\ Rev.\ Lett.\  {\bf 112} (2014) 112001
  % doi:10.1103/PhysRevLett.112.112001
  [arXiv:1312.1228 [hep-ph]].
  %% CITATION = doi:10.1103/PhysRevLett.112.112001;%%

\bibitem{Cirigliano:2011ny} V.~Cirigliano, G.~Ecker, H.~Neufeld, A.~Pich and J.~Portoles,
  %\emph{Kaon Decays in the Standard Model},
  Rev.\ Mod.\ Phys.\  {\bf 84} (2012) 399
  % doi:10.1103/RevModPhys.84.399
  [arXiv:1107.6001 [hep-ph]].
  %% CITATION = doi:10.1103/RevModPhys.84.399;%%
  % 108 citations counted in INSPIRE as of 16 Jul 2016

\bibitem{Kaneko:2012cta}
 T.~Kaneko {\it et al.}  [JLQCD Collaboration],
  %\emph{Chiral behavior of kaon semileptonic form factors in lattice QCD with exact chiral symmetry},
  PoS  {\bf LATTICE2012} (2012) 111
  [arXiv:1211.6180 [hep-lat]].
  %% CITATION = ARXIV:1211.6180;%%

\bibitem{Boyle:2013gsa}
 P.~A.~Boyle, J.~M.~Flynn, N.~Garron, A.~J\"uttner, C.~T.~Sachrajda, K.~Sivalingam, and J.~M.~Zanotti
  [RBC/UKQCD Collaboration],
  %\emph{The kaon semileptonic form factor with near physical domain wall quarks},
  JHEP {\bf 1308} (2013) 132
  [arXiv:1305.7217 [hep-lat]].
  %% CITATION = ARXIV:1305.7217;%%

\bibitem{KtopiLat2013}
 E.~G\'amiz {\it et al.}  [Fermilab Lattice and MILC Collaborations],
  %\emph{Kaon semileptonic form factors with $N_f=2+1+1$ HISQ fermions and physical light quark masses},
  PoS {\bf LATTICE2013} (2013) 395  [arXiv:1311.7264 [hep-lat]].
  %% CITATION = ARXIV:1311.7264;%%


\bibitem{Boyle:2015hfa}
 P.~A.~Boyle {\it et al.} [RBC/UKQCD Collaboration],
  %\emph{The kaon semileptonic form factor in N$_{f}$ = 2 + 1 domain wall lattice QCD with physical light quark masses},
  JHEP {\bf 1506} (2015) 164
  % doi:10.1007/JHEP06(2015)164
  [arXiv:1504.01692 [hep-lat]].
  %% CITATION = doi:10.1007/JHEP06(2015)164;%%

\bibitem{Carrasco:2016kpy}
 N.~Carrasco, P.~Lami, V.~Lubicz, L.~Riggio, S.~Simula and C.~Tarantino,
  %\emph{$K \to \pi$ semileptonic form factors with $N_f=2+1+1$ twisted mass fermions},
  Phys.\ Rev.\ D {\bf 93} (2016) 114512
  % doi:10.1103/PhysRevD.93.114512
  [arXiv:1602.04113 [hep-lat]].
  %% CITATION = doi:10.1103/PhysRevD.93.114512;%%

\bibitem{Aoki:2016frl}
 S.~Aoki {\it et al.},
  %\emph{Review of lattice results concerning low-energy particle physics},
  arXiv:1607.00299 [hep-lat].
  %% CITATION = ARXIV:1607.00299;%%

\bibitem{Gamiz:2016bpm}
  E.~Gámiz {\it et al.} [Fermilab Lattice and MILC Collaborations],
  %``Kaon semileptonic decays with $N_f=2+1+1$ HISQ fermions and physical light-quark masses,''
  arXiv:1611.04118 [hep-lat].
  %%CITATION = ARXIV:1611.04118;%%

\bibitem{Bazavov:2012cd}
 A.~Bazavov {\it et al.},
 %\emph{Kaon semileptonic vector form factor and determination of $|V_{us}|$ using staggered fermions},
  Phys.\ Rev.\ D {\bf 87} (2013) 073012
  % doi:10.1103/PhysRevD.87.073012
  [arXiv:1212.4993 [hep-lat]].
  %% CITATION = doi:10.1103/PhysRevD.87.073012;%%

\bibitem{Na:2009au}
 H.~Na, C.~T.~H.~Davies, E.~Follana, P.~Lepage and J.~Shigemitsu,
  %\emph{D semi-leptonic decay form factors with HISQ charm and light quarks},
  PoS LAT {\bf 2009} (2009) 247
  [arXiv:0910.3919 [hep-lat]].
  %% CITATION = ARXIV:0910.3919;%%
  % 8 citations counted in INSPIRE as of 16 Jul 2016

\bibitem{Koponen:2012di}
 J.~Koponen {\it et al.} [HPQCD Collaboration],
  %\emph{D to K and D to pi semileptonic form factors from Lattice QCD},
  arXiv:1208.6242 [hep-lat].
  %% CITATION = ARXIV:1208.6242;%%

\bibitem{Bernard:2013eya}
 C.~Bernard, J.~Bijnens and E.~Gámiz,
  %\emph{Semileptonic Kaon Decay in Staggered Chiral Perturbation Theory},
  Phys.\ Rev.\ D {\bf 89} (2014) 054510
  % doi:10.1103/PhysRevD.89.054510
  [arXiv:1311.7511 [hep-lat]].
  %% CITATION = doi:10.1103/PhysRevD.89.054510;%%

\bibitem{Ghorbani}
 K.~Ghorbani,
  %``Chiral and Volume Extrapolation of Pion and Kaon Electromagnetic form Factor within SU(3) ChPT,''
  Chin.\ J.\ Phys.\  {\bf 51} (2013) 920
  %doi:10.6122/CJP.51.903, 10.6122/CJP.51.920
  [arXiv:1112.0729 [hep-ph]].
  %%CITATION = doi:10.6122/CJP.51.903, 10.6122/CJP.51.920;%%

\bibitem{Bijnens:2014yya}
 J.~Bijnens and J.~Relefors,
 % \emph{Masses, Decay Constants and Electromagnetic Form-factors with Twisted Boundary Conditions},
  JHEP {\bf 1405} (2014) 015
  %% doi:10.1007/JHEP05(2014)015
  [arXiv:1402.1385 [hep-lat]].
  %% CITATION = doi:10.1007/JHEP05(2014)015;%%

\bibitem{Jiang:2006gna} F.-J.~Jiang and B.~C.~Tiburzi,
  %``Flavor twisted boundary conditions, pion momentum, and the pion electromagnetic form-factor,''
  Phys.\ Lett.\ B {\bf 645}, 314 (2007)
  %doi:10.1016/j.physletb.2006.12.041
  [hep-lat/0610103].
  %%CITATION = doi:10.1016/j.physletb.2006.12.041;%%

\bibitem{CHIRON}
 J.~Bijnens,
  %\emph{CHIRON: a package for ChPT numerical results at two loops},
  Eur.\ Phys.\ J.\ C {\bf 75} (2015) 27
  % doi:10.1140/epjc/s10052-014-3249-9
  [arXiv:1412.0887 [hep-ph]].
  %% CITATION = doi:10.1140/epjc/s10052-014-3249-9;%%

\bibitem{Bazavov:2012xda}
 A.~Bazavov {\it et al.} [MILC Collaboration],
  % ``Lattice QCD ensembles with four flavors of highly improved staggered quarks,''
  Phys.\ Rev.\ D {\bf 87} (2013) 054505
  %doi:10.1103/PhysRevD.87.054505
  [arXiv:1212.4768 [hep-lat]].
  %% CITATION = doi:10.1103/PhysRevD.87.054505;%%

\bibitem{thesisjohan} J. Relefors, ``Twisted Loops and Models for
Form-factors and the Muon $g-2$,'' PhD thesis, Lund University, September 2016,
ISBN 978-91-7623-975-9.

\bibitem{Weinberg} S.~Weinberg,
  %\emph{Phenomenological Lagrangians},
  Physica A {\bf 96} (1979) 327.
  %% CITATION = PHYSA,A96,327;%%

\bibitem{GL1} J.~Gasser and H.~Leutwyler,
  %\emph{Chiral Perturbation Theory To One Loop},
  Annals Phys.\  {\bf 158} (1984) 142;
  %% CITATION = APNYA,158,142;%%

\bibitem{GL2} J.~Gasser and H.~Leutwyler,
  %\emph{Chiral Perturbation Theory: Expansions In The Mass Of The Strange Quark},
  Nucl.\ Phys.\ B {\bf 250} (1985) 465.
  %% CITATION = NUPHA,B250,465;%%

\bibitem{GL3} J.~Gasser and H.~Leutwyler,
  %\emph{Low-Energy Expansion of Meson Form-Factors},
  Nucl.\ Phys.\ B {\bf 250} (1985) 517.
  %% CITATION = NUPHA,B250,517;%%

\bibitem{GL4} J.~Gasser and H.~Leutwyler,
  %\emph{Spontaneously Broken Symmetries: Effective Lagrangians At Finite Volume},
  Nucl.\ Phys.\ B {\bf 307} (1988) 763.
  %% CITATION = NUPHA,B307,763;%%

\bibitem{Sharpe:2001fh} S.~R.~Sharpe and N.~Shoresh,
  %\emph{Partially quenched chiral perturbation theory without Phi0},
  Phys.\ Rev.\ D {\bf 64} (2001) 114510
  % doi:10.1103/PhysRevD.64.114510
  [hep-lat/0108003].
  %% CITATION = doi:10.1103/PhysRevD.64.114510;%%

\bibitem{Sharpe:1992ft} S.~R.~Sharpe,
  %\emph{Quenched chiral logarithms},
  Phys.\ Rev.\ D {\bf 46} (1992) 3146
  % doi:10.1103/PhysRevD.46.3146
  [hep-lat/9205020].
  %% CITATION = doi:10.1103/PhysRevD.46.3146;%%

\bibitem{Bernard:1993sv}
  C.~W.~Bernard and M.~F.~L.~Golterman,
  %\emph{Partially quenched gauge theories and an application to staggered fermions},
  Phys.\ Rev.\ D {\bf 49} (1994) 486
  % doi:10.1103/PhysRevD.49.486
  [hep-lat/9306005].
  %% CITATION = doi:10.1103/PhysRevD.49.486;%%

\bibitem{Damgaard:2000gh}
  P.~H.~Damgaard and K.~Splittorff,
  %\emph{Partially quenched chiral perturbation theory and the replica method},
  Phys.\ Rev.\ D {\bf 62} (2000) 054509
  % doi:10.1103/PhysRevD.62.054509
  [hep-lat/0003017].
  %% CITATION = doi:10.1103/PhysRevD.62.054509;%%

\bibitem{Lee:1999zxa}
  W.~J.~Lee and S.~R.~Sharpe,
  %\emph{Partial flavor symmetry restoration for chiral staggered fermions},
  Phys.\ Rev.\ D {\bf 60} (1999) 114503
  % doi:10.1103/PhysRevD.60.114503
  [hep-lat/9905023].
  %% CITATION = doi:10.1103/PhysRevD.60.114503;%%

\bibitem{Aubin:2003mg}
  C.~Aubin and C.~Bernard,
  %\emph{Pion and kaon masses in staggered chiral perturbation theory},
  Phys.\ Rev.\ D {\bf 68} (2003) 034014
  % doi:10.1103/PhysRevD.68.034014
  [hep-lat/0304014];
  %% CITATION = doi:10.1103/PhysRevD.68.034014;%%
  C.~Aubin and C.~Bernard,
  %\emph{Pseudoscalar decay constants in staggered chiral perturbation theory},
  Phys.\ Rev.\ D {\bf 68} (2003) 074011
  % doi:10.1103/PhysRevD.68.074011
  [hep-lat/0306026].
  %% CITATION = doi:10.1103/PhysRevD.68.074011;%%

\bibitem{Bedaque}
  P.~F.~Bedaque,
  %\emph{Aharonov-Bohm effect and nucleon-nucleon phase shifts on the lattice},
  Phys. Lett. B {\bf 593} (2004) 82
  [arXiv:nucl-th/0402051]
  %% CITATION = NUCL-TH/0402051;%%

\bibitem{twisted}
 C.~T.~Sachrajda and G.~Villadoro,
  %\emph{Twisted boundary conditions in lattice simulations},
  Phys.\ Lett.\ B {\bf 609} (2005) 73
  [arXiv:hep-lat/0411033].
  %% CITATION = HEP-LAT/0411033;%%

\bibitem{Na:2010uf}
 H.~Na, C.~T.~H.~Davies, E.~Follana, G.~P.~Lepage and J.~Shigemitsu,
  %\emph{The $D \rightarrow K, l \nu$ Semileptonic Decay Scalar Form Factor and $|V_{cs}|$ from Lattice QCD},
  Phys.\ Rev.\ D {\bf 82} (2010) 114506
  % doi:10.1103/PhysRevD.82.114506
  [arXiv:1008.4562 [hep-lat]].
  %% CITATION = doi:10.1103/PhysRevD.82.114506;%%

\bibitem{Kaneko:2016mha}
 T.~Kaneko {\it et al.} [JLQCD Collaboration],
  %``Chiral behavior of light meson form factors in 2+1 flavor QCD with exact chiral symmetry,''
  PoS LATTICE {\bf 2015} (2016) 325
  [arXiv:1601.07658 [hep-lat]].
  %%CITATION = ARXIV:1601.07658;%%

\bibitem{Bijnens:2003uy}
  J.~Bijnens and P.~Talavera,
  %\emph{K(l3) decays in chiral perturbation theory},
  Nucl.\ Phys.\ B {\bf 669} (2003) 341
  [hep-ph/0303103].
  % doi:10.1016/S0550-3213(03)00581-9
  %% CITATION = doi:10.1016/S0550-3213(03)00581-9;%%

\bibitem{Bazavov:2011aa}
 A.~Bazavov {\it et al.} [Fermilab Lattice and MILC Collaborations],
  %``B- and D-meson decay constants from three-flavor lattice QCD,''
  Phys.\ Rev.\ D {\bf 85} (2012) 114506
  doi:10.1103/PhysRevD.85.114506
  [arXiv:1112.3051 [hep-lat]].
  %%CITATION = doi:10.1103/PhysRevD.85.114506;%%

\bibitem{Ktopilnu2016}
 A.~Bazavov {\it et al.} [Fermilab Lattice and MILC Collaborations], in preparation.

\bibitem{hairpins}
  A.~Bazavov {\it et al.}  [MILC Collaboration],
  % ``Properties of light pseudoscalars from lattice QCD with HISQ ensembles,''
  PoS LATTICE {\bf 2011} (2011) 107 [arXiv:1111.4314], and work in progress

\end{thebibliography}
\end{document}